\documentclass[nonacm,sigconf]{acmart}
\usepackage{listings}
\usepackage[utf8]{inputenc}
\usepackage{pmboxdraw}
\usepackage{multicol}
\usepackage{pifont}
\usepackage{graphicx}
\usepackage{dcolumn}
\usepackage{bm}
\usepackage[T1]{fontenc}
\usepackage{url}
\usepackage{fancyhdr}
\usepackage{verbatim}
\usepackage{bookmark}
\usepackage{changepage}
\usepackage{float}
\usepackage{amsmath}
\usepackage{tikz}
\usepackage{gincltex}
\usepackage[normalem]{ulem}
\usepackage{listings}
\usepackage[group-separator={,},binary-units=true]{siunitx}
\usepackage{multirow}
\usepackage{ifthen}

\AtBeginDocument{\renewcommand{\vec}[1]{\overrightarrow{#1}}}
\DeclareMathDelimiter{(}{\mathopen} {operators}{"28}{largesymbols}{"00}
\DeclareMathDelimiter{)}{\mathclose}{operators}{"29}{largesymbols}{"01}

\AtBeginDocument{%
  \providecommand\BibTeX{{%
    \normalfont B\kern-0.5em{\scshape i\kern-0.25em b}\kern-0.8em\TeX}}}

\setcopyright{acmcopyright}
\copyrightyear{2020}
\acmYear{2020}
\acmDOI{10.1145/1122445.1122456}

\acmConference[SIGMETRICS '20]{ACM SIGMETRICS 2020}{June 8--12, 2020}{Boston, MA}

\newcounter{tbdcounter}
\newcommand{\HL}[1]{ #1}

\newcommand{\isrc}[1]{\lstinline[basicstyle=\ttfamily]|#1|}

\DeclareSIUnit \terawatthour{{\tera\watt\hour}} 
\DeclareSIUnit{\million}{\text{million}}
\DeclareSIUnit{\billion}{\text{billion}}
\DeclareSIUnit{\USD}{\text{USD}}
\DeclareSIUnit{\ETH}{\text{ETH}}
\DeclareSIUnit{\BIGM}{\text{M}}

\newbool{INTRO}
\newbool{ARCHITECTURE}
\newbool{METHOD}
\newbool{RESULTS}
\newbool{COSTMODEL}
\booltrue{INTRO}        %
\booltrue{ARCHITECTURE} 
\booltrue{METHOD}       
\booltrue{RESULTS}      %
\booltrue{COSTMODEL}    %
\InputIfFileExists{config.tex}{\input{config.tex}}{}

\usepackage{hyperref}

\makeatletter
\newcommand{\printfnsymbol}[1]{%
  \textsuperscript{\@fnsymbol{#1}}%
}
\makeatother

\begin{document}

\title{The Economics of Smart Contracts}

\author{Kirk Baird*}
\affiliation{The University of Sydney} 
\email{kbai2800@uni.sydney.edu.au}
\author{Seongho Jeong*}
\affiliation{Yonsei University}
\email{seongho.jeong@yonsei.ac.kr}
\author{Yeonsoo Kim*}
\affiliation{Yonsei University}
\email{yeonsoo.kim@yonsei.ac.kr}
\author{Bernd Burgstaller}
\affiliation{Yonsei University}
\email{bburg@yonsei.ac.kr} 
\author{Bernhard Scholz}
\affiliation{The University of Sydney} 
\email{Bernhard.Scholz@sydney.edu.au}

\thanks{*Authors contributed equally.}
\begin{abstract}
Ethereum is a distributed blockchain that can execute
smart contracts, which inter-communicate and perform transactions automatically.
The execution of smart contracts is paid in the form of gas, which is a
monetary unit used in the Ethereum blockchain. The Ethereum Virtual
Machine~(EVM) provides the metering capability for smart contract execution.
Instruction costs vary depending on the instruction type and the approximate
computational resources required to execute the instruction on the network. The
cost of gas is adjusted using transaction fees to ensure adequate payment of
the network.

In this work, we highlight the ``real'' economics of smart contracts. We show
that the actual costs of executing smart contracts are disproportionate to the
computational costs and that this gap is continuously widening. We show that the gas
cost-model of the underlying EVM instruction-set is wrongly modeled.
Specifically, the computational cost for the \isrc{SLOAD}~instruction increases with
the length of the blockchain. Our proposed performance model estimates
gas usage and execution time of a smart contract at a given block-height.
The new gas-cost model incorporates the block-height to eliminate
irregularities in the Ethereum gas calculations.
Our findings are based on extensive
experiments over the entire history of the EVM blockchain.
\end{abstract}



\maketitle

\begin{section}{Introduction\label{sec:intro}}
\ifINTRO
Ethereum is the largest blockchain with the ability to execute
arbitrarily-expressive computations called smart contracts.
A smart contract can capture complex business transactions
involving various accounts by dispensing and accepting funds.
Developers commonly write smart contracts in a high-level language~\cite{solidity,serpent,vyper}
that is compiled to bytecode, deployed on the blockchain, and later
executed on the distributed Ethereum Virtual Machine~(EVM).

The business applications for smart contracts are manifold,
including prediction
markets, governance, investment organizations, crowdfunding, music
distribution, and many others~\cite{Rose17icos,Fortune:ICOs,ICOstats,pilkington201611}.
Smart contracts have been receiving attention from
economists, lawyers, the technology industry, and governments~\cite{Tapscott2016,Corda2017,Flood2015,UK2016}.
Smart contracts perform transactions in Ethereum's {\em Ether} cryptocurrency (ETH).
In 2019, Ethereum has a market capitalization of \SI{19}{\billion\USD}~\cite{etherscan-marketcap}.

In Ethereum, a user issues transactions to a peer-to-peer~(P2P) network of mutually distrusting
nodes, which employ miners to collate submitted transactions into blocks.
A consensus protocol
determines the next block for execution on the network.
Once a block is selected, each transaction in the block is processed.
A transaction consists of smart contract executions,
the possible transfer of Ether between accounts,
and the writing of the side-effects
to the shared global state on the blockchain.
A miner in the P2P network is incentivized with (1)~a block reward for
producing the next block through the consensus protocol, and (2)~transaction
fees for executing the transactions of this block.
Block rewards are minted and issued by the network as an incentive for
miners to operate and secure the network. Transaction fees are paid
by the users issuing transactions.

The Ethereum consensus protocol is currently in a tran\-si\-tion-phase from
proof-of-work (PoW~\cite{Jakobsson:1999:POW}) to proof-of-stake
(PoS~\cite{Buterin2017}), to eliminate the high computational cost incurred by
PoW-mining~\cite{bitc:carb}.
With PoS, miners (then called \emph{validators}\footnote{We
use the term \emph{miner} to refer to both PoW-miners and PoS-validators
where the distinction is clear from context.}) will
deposit a minimum of \SI{32}{\ETH} to become partakers in the consensus
mechanism.  Miners are selected pseudo-randomly by the consensus protocol
to produce blocks~\cite{Ethereum:pos:cost}. The block rewards from
participation in the PoS consensus protocol resemble an annual interest rate on
the staked deposit~\cite{Casper:incentives}.  The projected PoS block rewards
are considerably below current PoW rates, to cut the inflation from
cryptocurrency minted to pay rewards~\cite{Ethereum:pos:issuance}.  In
contrast, the transaction fees for block execution will carry over to the PoS
consensus protocol. As with PoW, transaction fees will be paid to the
miner who
created the block~\cite{Casper:incentives,Ethereum:pos:transaction:fees,Ethereum:pos:cost}.

As soon as the much lighter PoS will have replaced Ethereum's computationally
intense PoW consensus protocol, the computations from smart contract execution
will dominate the total runtime.
Due to the reduced block rewards from PoS, transaction
fees will become a significant source of income for miners.
For the new PoS protocol,
it will be critical whether
\begin{enumerate}
\item \emph{transaction fees are proportional to the computational costs} (CPU
costs, energy costs) that miners have to spend for executing smart
contracts, and
\item  \emph{the charged fees are
adequate for the provided computational services from the users' point of view}.
\end{enumerate}
\HL{In this paper, our focus is exclusively on \emph{transaction fees\/}---the monetary value
paid by Ethereum users to Ethereum miners for executing transactions
on the blockchain. As outlined above, the cost model for
transaction fees is oblivious to the consensus protocol, and our discussion
thus applies to both the current (Pow) and upcoming (PoS) version of the
Ethereum blockchain.  Despite the value represented by block rewards, we argue
that the cost model for transaction fees is important in itself, because
(1)~transactions constitute a major source of income
in the upcoming PoS consensus protocol, and (2)~an accurate cost model is crucial to secure the
dependable and scalable operation of the Ethereum blockchain.
In particular, the Ethereum blockchain already fell victim to denial-of-service (DoS)
attacks that exploited under-priced operations in smart contracts to slow down
the processing of blocks~\cite{chen2017adaptive}.}

\begin{figure}[t]
  \hspace{-2.5mm} 
  \includegraphics[width=80mm]{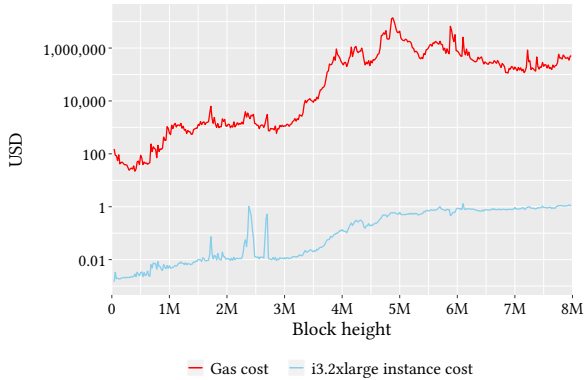}
  \caption{Comparison between transaction fees (gas cost) and \HL{transaction execution costs on AWS.}}
  \label{fig:gas-vs-aws}
\end{figure}

Computations on the Ethereum blockchain are metered in units of \emph{gas}.
Users specify upfront the \emph{GasLimit}, which is the maximum amount of gas
committed to the transaction.  Transferring Ether between accounts consumes a
minimum amount of \num{21000} units of gas. This amount increases with account
creation and for the execution of a smart contract on the EVM.  Each EVM
instruction costs a certain amount of gas, as defined in the Ethereum Yellow
Paper~\cite{Wood2014}.  During execution, the EVM meters the gas cost per
executed instruction. If the gas usage exceeds the transaction's GasLimit, the EVM will halt
the transaction \HL{(but nevertheless charge the consumed gas to the user)}, which is a mechanism to prevent unbounded, potentially non-terminating
computations and resource abuse over
the network. The gas costs of transactions are charged to users
in units of Ether. A transaction contains a \emph{GasPrice}, which
is the amount of Ether the user is willing to pay per unit of gas.
The consumed units of gas times the GasPrice is the fee
paid by the transaction.
Users can set the GasPrice of a transaction deliberately higher to prioritize a
transaction in the pool of submitted transactions. Because miners want
to maximize revenue, transactions with a high GasPrice will be selected first
for execution on the blockchain.

This work is an empirical investigation to find out the extent to which transaction fees match the
costs spent by miners to execute transactions.
For this purpose, we instrumented the Ethereum Parity client~\cite{ParityWiki}
to obtain execution time and gas cost of each executed block.
With the instrumented client, we processed the Ethereum
blockchain from the genesis block up to Block~\SI{8}{\BIGM}
(originally mined on 21/06/2019).
This experiment was conducted
on an Amazon AWS i3.2xlarge instance. The instance
provides an NVMe SSD to facilitate low-latency database access and thus
constitutes an obvious choice for a PoS-mining service on a rented Cloud node.
To compare Ethereum and AWS fees,
we converted transaction fees from Ether to USD based on historical
Ethereum price data~\cite{etherscan-etherprice}.

\HL{The results of our investigation are depicted in Figure~\ref{fig:gas-vs-aws} (please note
again that our focus is on transaction fees and thus neither PoW-mining hardware costs
which will be eliminated with PoS, nor block rewards which will be
diminished with PoS, are included in this diagram.)} 
We observed a considerable
disparity between the transaction fee paid by a user and the cost
of the miner. From block \num{4880000} to
\num{4900000}, the value of the Ether cryptocurrency was at its peak, and the sum of
the transaction fees within this range was the equivalent of \SI{14}{\million\USD}, whereas
the actual cost of computation on AWS was less than \SI{1}{\USD}.
This huge gap would not close even if every single node
in the entire Ethereum network would be paid transaction fees (e.g.,
on 09/30/2019 the Ethereum network contained \num{7283}~nodes~\cite{etherscan-nodetracker},
and the historically largest network-size the authors are aware of are \num{15454}~nodes
on 04/23/2018, as reported in~\cite{kim2018measuring}).

\begin{figure}[t]
  \centering
  \includegraphics[width=80mm]{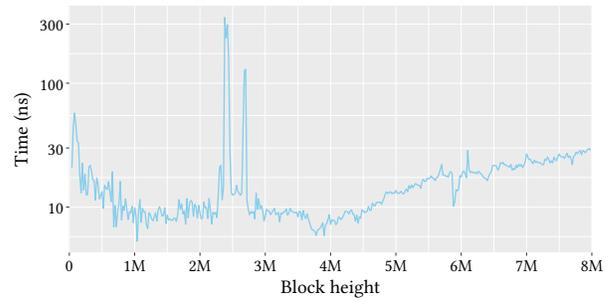}
  \caption{Average \HL{transaction} execution time spent per unit gas.}
  \label{fig:time-per-gas}
\end{figure}

Another irregularity that surfaced in our experiments is the average execution
time spent by a miner to earn one unit of gas throughout the entire
blockchain. As depicted in Figure~\ref{fig:time-per-gas}, the execution time
per unit of gas has been steadily increasing after block~\SI{3.8}{\BIGM}.
(The outliers between
blocks \SI{2}{\BIGM} and \SI{3}{\BIGM} are due to the before-mentioned DoS attacks,
which abused the gas pricing of certain EVM instructions to significantly
slow down the processing of blocks~\cite{chen2017adaptive}.)
The trend of increasing execution time per unit of gas puts
miners at a disadvantage: to earn an equal amount of gas, miners need to use more
computing power over time.
Related findings have been recently reported in the Ethereum community,
indicating that CPU resources are not properly aligned with gas
costs~\cite{geth-measure}.

\HL{
We found that the Ethereum cost model does not accurately reflect the resource
costs incurred by smart contracts.
  From an economic point of view, the cost per
transaction should be proportional to the consumed gas.
However, this is currently not the case. For certain EVM instructions, the cost
model inaccurately predicts the resource consumption. In particular, we observed
that the current cost model is unable to reflect fluctuations that are
dependent on the block height (as suggested by Figure~\ref{fig:time-per-gas}).
Such inaccuracies make the Ethereum network susceptible to attacks that utilize
a vast amount of resources for a small cost. A new wave of DoS attacks is one threat, as
already voiced in the recent research literature~\cite{perez2019broken}.
Malicious compilers may exploit such inaccuracies in bytecode that disadvantages
miners and other users who are unaware of the cost-model's weakness.
Computational costs that increase with the length of the blockchain will
affect performance and scalability, as already articulated in the database
community~\cite{BC:DB,caper}.}
Because a block reward in the PoW consensus
protocol is significantly higher than the associated transaction fees, the cost
model for transactions is currently less relevant to miners.  However, once the
PoW protocol and its associated block reward are replaced by the energy-efficient PoS protocol,
transaction fees will become a more significant income source for miners.
Hence, a transparent and accurate cost model is paramount
for establishing a dependable, sustainable and  secure economy in
Ethereum. To establish such a cost model, this paper makes
the following contributions:

\begin{figure*}[t]
  \centering
   \includegraphics[width=124mm]{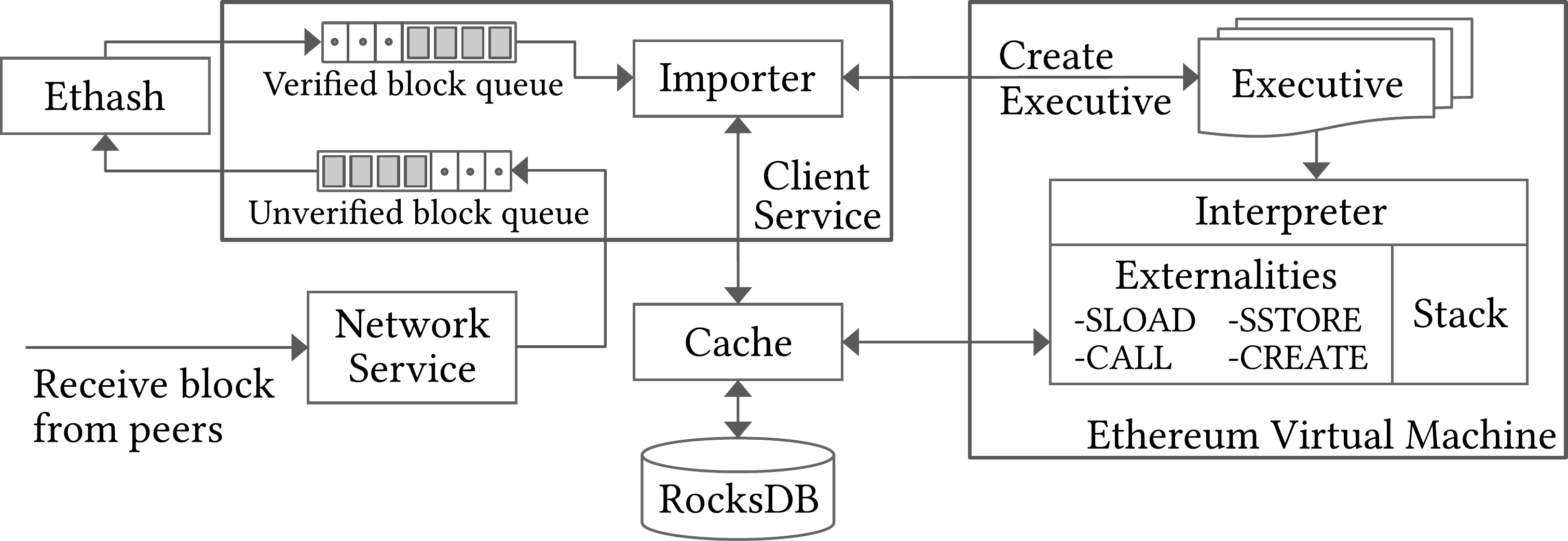}
  \caption{Conceptual view of Parity Ethereum client software architecture.}
  \label{fig:parity-abstraction}
\end{figure*}

\begin{enumerate}
  \item We provide a measurement scheme to determine the key performance characteristics of
     smart contracts. 
     Performance data is obtained and evaluated on both a macroscopic (func\-tion\-al\-i\-ty-based)
     and a microscopic (EVM in\-struc\-tion-level) view.
  \item We obtain the performance profiling results from the Parity Ethereum client
        for the entire \SI{8}{\BIGM}~blocks of the Ethereum blockchain.
  \item We provide evidence that the economics of the Ethereum gas cost model failed to reflect
       the actual computational costs. 
  \item We identify block-height as key for modeling the performance of smart contracts.
        We provide a performance model to estimate gas usage and execution time of a smart contract
        at a given block-height.
  \item We propose a new gas cost model that fixes the main irregularity of the current Ethereum gas cost
        model, i.e., the ongoing inflation of execution time per unit of gas.
\end{enumerate}

The rest of this paper is organized as follows.  In
Section~\ref{parity:arch}, we present the software architecture and
configuration options of the Parity Ethereum client.  Our performance
measurement scheme is introduced in Section~\ref{sec:measurements}.  We discuss the
data from our experimental evaluation in Section~\ref{sec:experiments}.  In
Section~\ref{gas-model}, we propose a new gas cost model based on our observations.
We discuss the related work in Section~\ref{sec:relwork} and draw our
conclusions in Section~\ref{sec:concl}.
\else 
  \label{fig:parity-abstraction}
  \label{fig:gas-vs-aws}
\fi 
\end{section}

\begin{section}{Parity Software Architecture\label{parity:arch}}
\ifARCHITECTURE

Ethereum as a service has core responsibilities that clients are required to fulfill:
\begin{itemize}
  \item P2P networking: propagation of new transactions and blocks to other clients.
  \item Block processing:
  checking the validity of a new block before executing the block's transactions and
  committing changes to the database.
  \item Shared global state (database) maintenance:
  storing the state trie which consists of all account balances, code that has been uploaded
  to the network (smart contracts), the current value of variables (storage
  slots) associated with each contract, and the blocks themselves.
\end{itemize}
Running an Ethereum client requires a connection to the Ethereum network to use
Ethereum boot nodes and the Ethereum peer discovery mechanism. Processing of
new blocks includes storing of balances, contract bytecode, contract variables,
and the blocks themselves. For clients to maintain the state of the network,
they must process every transaction starting from the genesis block, or
otherwise download a previous state, i.e., a snapshot, and process every
transaction after that point.

A range of clients have been developed for the Ethereum block\-chain.  The two
most popular clients are Geth~\cite{Geth} and Parity~\cite{ParityWiki}, with a
market share of \num{69.9}\% and \num{27.5}\%,
respectively~\cite{etherscan-nodetracker}.  The Parity implementation employs
the Rust programming language, whereas Geth is programmed in Go. Despite their
different implementation languages, we found the software architecture of
Parity and Geth to be virtually identical. We attribute this to the tight grip
that the Ethereum specification exercises over client implementations. For this paper, we
employ Parity, which has traditionally been known for its superior
performance~\cite{Ethereum-Benchmarks}.  In Section~\ref{sec:geth:arch}, we
show that our measurement-method carries over one-to-one to the software
architecture of Geth, modulo the renaming of functions.

\subsection{Core Components of the Parity Ethereum Client}
An overview of the software architecture of the Parity Ethereum client is
depicted in Figure~\ref{fig:parity-abstraction}.
Our analysis is based on Parity version~2.4.0~\cite{ParityGithub}.

\subsubsection*{Client Service}
The Client Service is the core service and constitutes a major part of the
system.  It handles peer requests, processes new blocks received over the P2P
network, and interacts with the database.  The Client Service maintains an
unverified and a verified block queue to process new blocks.  The
unverified block queue buffers new blocks as they are received over the network.  After the
check of signatures and block hashes, new blocks are moved to the verified block
queue.  Verified blocks are processed by the Importer, which will interact with
the database and the EVM as required.

\subsubsection*{Ethash}
Ethash verifies the block hashes and transaction signatures of the blocks
from the Client Service's unverified block queue.
The block hash, a keccak256 hash~\cite{bertoni2013keccak}, is verified
against the difficulty set by the PoW consensus algorithm.
Block verification requires the validation of ECSDA 256~bit transaction signatures.
Valid blocks are queued for block-import in the verified
block queue of the Client Service.

\subsubsection*{Ethereum Virtual Machine (EVM)}
The purpose of the EVM is to facilitate smart contract execution on the
Ethereum blockchain. Execution of a smart contract is often referred as
\emph{state execution}, because the computations of the smart contract
represent a transition from the shared global state prior to contract execution
to the shared global state after contract execution.  Smart contracts are
represented in the form of bytecode, which consists of opcodes such as
\isrc{ADD}, \isrc{POP}, \isrc{SLOAD}, and operands\footnote{EVM bytecode is
different from Java bytecode.  A discussion on the differences is provided
in~\cite{icse19}.}.  Certain EVM opcodes require interaction with the database
or other external resources, and these interactions are called externalities.
The EVM uses volatile memory for its execution, including a stack and main
memory.  Each call to a smart contract will create a new EVM instance with its
own stack/main memory and bytecode instructions, called an \emph{executive}.

\subsubsection*{Database}
Ethereum employs a state database to maintain states of accounts, transactions, and blocks.
Clients are required to maintain a local copy of the shared global state.
Parity uses RocksDB~\cite{RocksDB} as its underlying key-value database backend.

The Ethereum state database employs key-value pairs in the format $\langle\text{path}, \text{value}\rangle$.
Ethereum uses the recursive length prefix (RLP) encoding scheme~\cite{RLP}
 to encode data of arbitrary length. 
Pairs $\langle\text{path}, \text{value}\rangle$ are mapped to Merkel Patricia Tries~\cite{Patricia-Tree}.
Each node in a trie contains its value and the keys of its children (if child-nodes exist), i.e.,
$\text{node}=\langle\text{child}_{0}, \ldots,\text{child}_{n}, \text{value}\rangle$.
Lookups in the state-trie are not flat (unlike a key-value database): to look up
the value of a given key, the byte-representation of the key is converted to a \emph{path\/} (essentially
the sequence of nibbles of the key's byte-representation). 
To determine the trie's value for a given path, 
the first element of the sequence is used as the index into the root of the trie (which
determines the respective child-node).
The second element indexes the child-node, also, until after $n$ lookups (in the general case)
the value in the trie's leaves is reached.

Trie nodes themselves are stored as $\langle\text{key}, \text{node}\rangle$ tuples in
the  key-value database backend, i.e., RocksDB.
The key of a node is the same as the node if its data-representation is less than \SI{32}{bytes} wide,
otherwise the \SI{32}{byte}~hash value of the node becomes the key.
Because hash values of nodes are used as references to the nodes, when there is a change in a node,
its parent nodes must be updated to reflect the corrected hash values.
This property guarantees deterministic hash values of root nodes, thus clients only need to know
root hash values to validate or revert their state database.

\subsubsection*{Database Caches}
There are three different types of caches used by the Parity client in
addition to a memory overlay. (1)~The database cache is the internal cache used by 
RocksDB and thus outside the control of the client. The default memory allocated is
\SI{128}{MB} or 70\% of the configured cache size. (2)~The state cache is programmed by Parity
to store details about information associated with an account address such as balance, contract bytecode,
and storage slot values. The state cache is set to a default of \SI{25}{MB} or 20\% of the
configured cache size. (3)~A blockchain cache is employed to store information about blocks
such as the block header and transactions. The default size for the blockchain cache is \SI{8}{MB} or
10\% of the configured cache size.
The memory overlay is used by Parity around RocksDB to reduce the number of reads and writes.
The memory overlay works by storing the values of a certain number of blocks (default: 64) or
up to the given storage size (default: \SI{32}{MB}) in memory. 
%

\subsubsection*{Network Service}
The Network Service propagates new blocks and transactions through
the P2P network. The Network Service uses a discovery protocol to
find new peers before establishing a TCP connection.
Once a new block has been received, it is queued for verification in the 
unverified block queue of the Client Service.

\subsection{Modes of Operations and Configurations}
\subsubsection*{Modes}
The purpose of a \textit{full node} is to fulfill all the requirements of a client.
A full node is required to process all new transactions and blocks.
Subsequently, the full node propagates the new transactions
to other nodes on the network. The role of a full node demands
that it maintains the entirety of the shared global state.
In contrast, a \textit{light node} is a type of Ethereum client that does not store the entire state
of the blockchain nor completely processes blocks/transactions. The purpose of a light
client is to interact with full nodes on the network in order to push
transactions or retrieve data only when required.

\subsubsection*{Pruning}
The \textit{archive pruning} mode stores the state of the blockchain at every block-height from genesis to present.
It is generally not advisable for most users to run it as it dramatically increases
space requirements. The typical users are blockchain explorers (websites that
display information about the chain) and users wishing to do analysis on the
state of the chain.
\textit{OverlayRecent (fast) pruning} separates blocks into ancient and recent
with only a certain number of
recent blocks being maintained. The difference is that recent blocks are
stored in memory and flushed at the end of the recent period. This is the
default setting for a client and the setting used in our analysis.
\fi 
\end{section}

\begin{section}{Measurement Methodology\label{sec:measurements}}
\ifMETHOD
We manually instrumented the Parity code with time measurement routines to measure 
the performance of particular sections of the code.
Mutex locks were 
required in some components because of the multi-threaded nature of Parity. Attempts
were made to ensure that the impact of the instrumentation on performance was kept to a
minimum.

\subsubsection*{Parity Configuration}
Unless otherwise specified, the Parity client was run with the default configuration,
which includes a default cache of \SI{169}{\mega\byte} and further \SI{32}{\mega\byte} for pruning
memory.

\subsubsection*{Importing Blocks from File}
In our initial profiling runs, we configured Parity to use the P2P network for
obtaining the block history (a profiling run essentially constitutes a re-play
of the block history of a certain range of blocks). The obtained performance
data showed variations of over 100\% for the initial one million blocks (for
otherwise identical profiling runs).  These variances occurred due to bandwidth
variations on the Internet, and from changes in the willingness of peers to
send our client the history of blocks (exporting historical blocks to peers is
a non-priority task with Ethereum clients).  As a result, we stopped obtaining
blocks from the P2P network in our experiments.

Instead, we obtained the entire history of blocks by running our own client
from the genesis block until the head of the chain. From this client, we exported all blocks
onto a local disk.  Because of the potential bottleneck of disk~I/O, we stored
the exported blocks on a dedicated disk different from the disk that contained
the Parity database, to minimize the performance impact of reading the blocks
from disk.

Thus for all reported experimental results, block import was conducted from a local
disk, to avoid network conditions to perturb the experiments.

\subsubsection*{Parallelism of Verification}
As mentioned before, our profiling experiments
were run post-facto on the block history, which implies that all blocks were
available at all times. Although processing transactions is a
serialized process, verifying blocks can be done in parallel.
Parallel verification is less frequent for fully synced (online) clients, as one canonical block is
produced approximately every \SI{15}{\s}.
Hence, only one or a small number of
blocks will be available for verification at any one time. The impact may increase the
total number of blocks processed per second compared to if blocks were processed
one at a time. However, because we aggregate time on a per-thread basis, the total
amount of time spent verifying blocks and importing blocks is not affected
by parallel block verification.

\subsubsection*{Limitations of Wall Clock Time}
A limitation of the manual instrumentation is that it uses wall clock time
to measure resource constraints. Wall clock time can be overstated as it does not
account for threads being paused by the kernel nor time spent waiting for locks.
Nevertheless, the hardware platform used for our experiments (see Table~\ref{table:HWspec})
contained a total number of \num{44}~cores 
to ensure that cores were not
over-subscribed. 

\subsection{Macroscopic Instrumentation}
Our macroscopic instrumentation examines Parity from a high-level view (on the level
of functions and components) to determine 
where the bottlenecks occur with block processing.

\subsubsection*{Logging Infrastructure}
The underlying logging infrastructure was implemented in C++ to minimize
performance overhead from the instrumentation.
Each log has a category, start time and 
stop time. Categories are used to segment the code into
different functional parts such as EVM and Block Verification. The results of
logs are stored as aggregate logs, which aggregate the duration (stop time - start time)
for all logs in the same category. As some categories may run  in parallel, a
mutex lock was required to protect the aggregated values from race conditions.
Each set of aggregate logs was archived every \num{20000}~blocks. The aggregation
window of \num{20000}~blocks is large enough to minimize daily fluctuations without
being too small to allow for granular analysis of trends.

\subsubsection*{Selection of Instrumentation Points}
The logging infrastructure required start and stop functions to be called from within
the Parity code. The points were chosen to analyze the main two categories of processing
blocks, i.e., block verification and block import.
Blocks can be verified in parallel. Importing, however is a serialized process because 
each transaction is dependent on the state of the previous transaction. As
a result, there exists a mutex lock within Parity's Importer. Within this lock, we inserted a
range of timers to further dissect block imports.

\subsubsection*{Gas Analysis}
The measurements of gas usage and general network statistics were taken by
manually injecting code into the Parity client to aggregate metrics
at a window-size of \num{20000}~blocks. Performance data was obtained by exporting the initial 8~million
blocks from our synced client. The blocks were then re-imported using the instrumented code.
Measurements were only taken from blocks that were on Ethereum's
canonical chain.

\subsection{Microscopic Instrumentation}
\subsubsection*{Static and Dynamic Behavior of EVM}
Our microscopic instrumentation of the Parity EVM interpreter was conducted
to understand both static and dynamic behavior of smart contract execution.
The \textit{static behavior} of contract execution on EVM
has to be identical for different invocations of contract code on the same state,
whereas the \textit{dynamic behavior} may differ across invocations.
E.g., gas consumption and world state encoded as a trie
are always identical across different client nodes on the network and are therefore
considered as static behavior.
The EVM consumes gas and it transitions between states by executing contracts. This state transition
is the static behavior defined by the Ethereum consensus protocol. It guarantees the integrity
of the smart contract execution over the network.

On the other hand, the performance of EVM and its underlying key-value database
to store the state trie constitutes dynamic behavior.
Dynamic behavior is expected
to vary across machines, implementations,
client modes and configurations.
In the following section, we focus on gas (i.e., static behavior) and execution time (i.e., dynamic
behavior).

\subsubsection*{Microscopic Instrumentation Granularity}
We conducted the microscopic instrumentation on two levels of granularity: on the granularity
of transactions and instructions.
The Ethereum blockchain only encodes the GasLimit and the GasPrice (see Section~\ref{sec:intro}) of a transaction.
To obtain the gas usage of each transaction, we manually instrumented Parity.
Gas usage and gas price of transactions are used in the calculation of the total amount of Ether paid
for each transaction in Figure~\ref{fig:gas-vs-aws}.
With our measurement of gas usage of EVM instructions, we did not
include the \num{21000} units of gas that Ethereum charges per transaction (the ``intrinsic'' gas). Rather, we measured
the net gas usage of the EVM interpreter instructions.
For instructions with non-constant gas cost functions, we measured all gas spent by the instructions
except gas provided for message calls of \isrc{CALL}, \isrc{CREATE}, and their derivations.
Gas and execution time of instructions are used in Section~\ref{gas-model} to
create our proposed gas cost model and to compare it to the existing gas cost model of Ethereum.

\subsubsection*{Low-level Synchronization Using Atomic Read-modify-write Operations}
The microscopic and macroscopic instrumentations had to be deployed in
separate profiling runs, because the logging infrastructure of the macroscopic
instrumentation imposes a too high overhead for the fine-grained performance
measurements of individual EVM instructions that we obtain from the microscopic
instrumentation.  For the aggregate counters of the
microscopic instrumentation, we employed atomic read-modify-write
operations~\cite{Boehm2008} to avoid the overhead of mutex locks.  At
transaction-level, the amount of consumed gas and the gas price of each transaction
was measured and aggregated per block.
At instruction-level, consumed gas, the execution time of each successful
instruction, and instruction call counts were aggregated at a window-size of
\num{1000}~blocks.
\fi 
\end{section}

\begin{section}{Experimental Results\label{sec:experiments}}
\ifRESULTS
We perform large-scale experiments over the entire Ethereum block\-chain in segments
of one million blocks 
using both the macroscopic and the microscopic instrumentation.
The macroscopic instrumentation is used to reveal the performance behavior of processing blocks,
which has two major categories, i.e., verify and import. The microscopic instrumentation is used
to analyze the fine-grained execution behavior of smart contracts.  To validate the macroscopic and microscopic
instrumentations, we compare their aggregated performance results against each other. For the experiment,
we use the computer hardware depicted in Table~\ref{table:HWspec}.

\begin{table}
   \small
  \caption{Hardware specification.\label{table:HWspec}}
    \begin{tabular}{|c|c|c|}
       \hline
       \multirow{4}{*}{CPU} & Model & Xeon E5-2699 v4\\
        & Manufacturer & Intel \\
        & Frequency & 2.2\textendash\SI{3.6}{\giga\hertz} \\
        & Cores (sockets) & 44 (2) \\
       \hline
       \multirow{5}{*}{Disk} & Type & SSD \\
        & Model & Optane 900P\\
        & Manufacturer & Intel\\
        & Interface & NVMe PCIe\\
        & Size & \SI{480}{\giga\byte} \\
       \hline
       \multirow{3}{*}{RAM} & Size &\SI{512}{\giga\byte} \\
        & Type & DDR4 \\
        & Speed & \SI{2400}{\mega\hertz}\\
       \hline
    \end{tabular}
\end{table}
\subsection{Block Processing\label{sec:block:proc}}
Block processing in the Parity client can be separated into two major categories,
verify and import. The two key features of verification are the validation of
transaction signatures and checking whether block hashes match the required difficulty.
Block import involves the processing of all transactions within a block and updating
the state trie of the database. For the block processing experiment, we
measured the duration of the following tasks.
\begin{itemize}
   \item Total time: measured from the start of a segment until the end, with segment sizes of \SI{1}{\million} blocks.
  \item Verify Block: verifying whether blocks are valid.
  \item Import Blocks: the serialized part of importing blocks where
  all transactions are processed and the results written to the database.
  \item Database commit (DB): a subset of Import Blocks; commits cached account changes to the database 
  after executing a transaction.
  \item Execute transaction (TX): a subset of Import Blocks that processes each
  transaction within the block.
  \item EVM: a subset of Executing Transactions; a measure of time spent directly inside the interpreter excluding set-up time of an EVM instance.
\end{itemize}
Table~\ref{table:block-processing-plang9-ssd} shows an aggregated overview of our experimental data. Values represent time in seconds and block ranges in millions.
\begin{table}[H]
\small
   \caption{Block processing time (s).\label{table:block-processing-plang9-ssd}}
  \centering
  \resizebox{\columnwidth}{!} {
    \begin{tabular}{c|cccccc}
    Blocks (M) & Total & Verify & Import & DB & TX & EVM \\
    \hline
    0 -- 1 & 708    & 4335  & 704    & 261   & 200    & 151   \\
    1 -- 2 & 2453   & 6013  & 2448   & 1026  & 966    & 843   \\
    2 -- 3 & 19535  & 6560  & 19529  & 4719  & 13651  & 10946 \\
    3 -- 4 & 9518   & 10850 & 9512   & 4630  & 3558   & 2112  \\
    4 -- 5 & 77038  & 36014 & 77032  & 28617 & 41453  & 25304 \\
    5 -- 6 & 157658 & 38232 & 157650 & 54754 & 94014  & 65926 \\
    6 -- 7 & 194231 & 30073 & 194224 & 61769 & 124277 & 96030 \\
    7 -- 8 & 239668 & 35023 & 239662 & 72958 & 157525 & 127277\\
    \end{tabular}
  }
\end{table}
A significant trend to note is the relationship between total time and block-height,
taking the time divided by the number of blocks processed yields time per block.
The measurements in Figure~\ref{fig:time-gas-per-block} illustrate that the time per block increases
as block-height increases, with outliers between blocks~\SI{2}{\BIGM} and \SI{3}{\BIGM}. The outliers are
due to the Shanghai DoS~\cite{chen2017adaptive} attacks, which caused a substantial increase in execution time.
Excluding the DoS attack period, the results loosely correlate the increasing time per block to gas per block ratio. 
As gas used per block increases, there is more strain on the client's resources, including CPU time and disk I/O, thereby reducing block processing speeds. However, the fact that there is only a loose correlation hints at possible discrepancies between gas costs and execution times.

\begin{figure}[h]
  \centering
  \includegraphics[width=80mm]{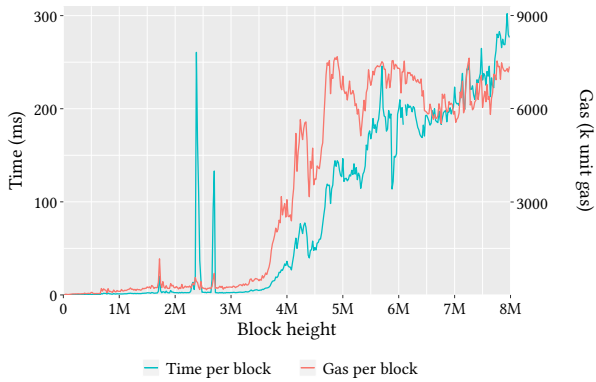}
   \caption{Time \& gas per block.}
  \label{fig:time-gas-per-block}
\end{figure}

The verification time increases with block-height in a pattern
almost identical to the number of transactions in a block as 
demonstrated in Figure~\ref{fig:block-verification-plang9-ssd}.
The cause of this relationship is heavily weighted on the number 
of ECDSA signatures that are processed in addition to verifying the block hash. 
Each transaction has a signature that is verified with elliptic curve multiplication operations
in $O\left(\log n\right)$ worst-case execution time. The validation of transaction signatures
accounts for the majority of verification time. Hence, we see a tight 
coupling between the number of transactions and the verification time.

\begin{figure}[h]
  \centering
  \includegraphics[width=80mm]{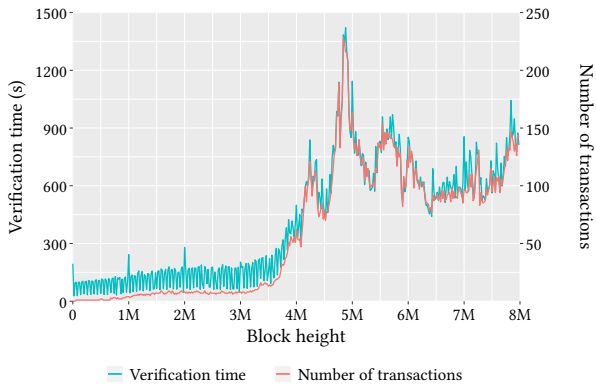}
  \caption{Verification time versus number of transactions.}
  \label{fig:block-verification-plang9-ssd}
\end{figure}

During the first \SI{3}{\million} blocks, the block-verify takes longer than block-import. However, the proportion of block-import  increases at a far greater rate as the block-height increases. Figure~\ref{fig:block-verify-import-plang9-ssd}
demonstrates the relationship between block-import and block-verify with respect to the block-height.
As block-import is growing at a higher rate than block-verify, we analyzed
block-import in more detail. The experiment for block-import is shown in Table~\ref{table:block-processing-plang9-ssd}, which implies that the proportion of block-import in total time 
is always almost 100\%. Because block-import is a serialized process taking 100\% 
of execution time, it can be concluded that it causes a performance bottleneck with 
block processing.



\begin{figure}[h]
  \centering
  \includegraphics[width=80mm]{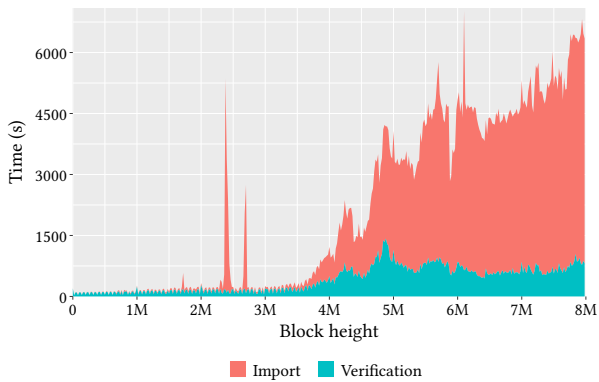}
  \caption{Block import vs. verification.}
  \label{fig:block-verify-import-plang9-ssd}
\end{figure}

\begin{figure}[h]
  \centering
  \includegraphics[width=80mm]{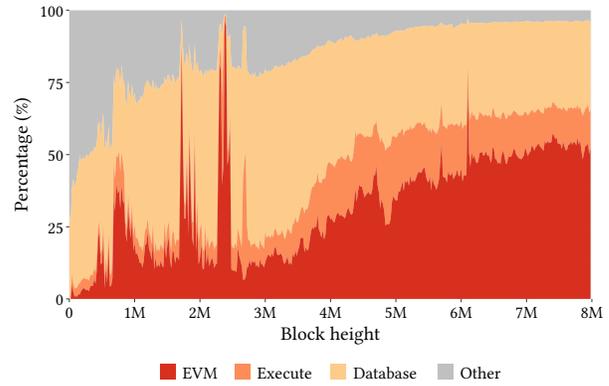}
  \caption{Block import split.}
  \label{fig:block-importation-plang9-ssd}
\end{figure}

\newpage
We further refined the analysis of block-import as shown in Figure~\ref{fig:block-importation-plang9-ssd}.
During the first \SI{3}{\million} blocks, there are large variations attributed to the
immaturity of the network and the Shanghai DoS attack. As the network continues to
mature, the block-import becomes more consistent with an increasing amount of
time being spent inside the EVM, reaching over 50\% at around block \SI{7}{\BIGM}.
This increase in EVM time justified the introduction of microscopic instrumentation for EVM 
instructions. However, there was still a significant amount of time 
spent in committing transactions to the database, which led to the 
performance analysis of caching.


The manual instrumentation was re-run with a cache-size 
set to \SI{32}{\giga\byte}. We cleared the cache every one
million blocks. Table~\ref{table:block-processing-plang9-ssd-cached} represents the $\frac{\text{time of default settings}}{\text{time of cached run}}$.
The result of the enlarged cache shows that the total time 
is 81\% faster. As the Parity client has caching at three levels, 
i.e., RocksDB, state, and blockchain, the results are relatively
 consistent between the categories of Import Blocks. 
The consistency between the categories implies there is a 
consistent portion of time spent on database operations.

\begin{table}[H]
\small
   \caption{Block processing time ratio --- 32 GB cache.\label{table:block-processing-plang9-ssd-cached}}
  \centering
  \resizebox{\columnwidth}{!} {
    \begin{tabular}{c|cccccc}
      Blocks (M) & Total & Verify & Import & DB & TX & EVM \\
      \hline
      0 -- 1 & 105\% & 100\% & 106\% & 108\% & 118\% & 123\% \\
      1 -- 2 & 112\% &  99\% & 112\% & 122\% & 121\% & 120\% \\
      2 -- 3 & 241\% & 101\% & 241\% & 142\% & 413\% & 597\% \\
      3 -- 4 & 130\% & 108\% & 130\% & 118\% & 215\% & 182\% \\
      4 -- 5 & 185\% & 100\% & 185\% & 165\% & 250\% & 214\% \\
      5 -- 6 & 230\% &  97\% & 230\% & 237\% & 267\% & 231\% \\
      6 -- 7 & 177\% &  97\% & 177\% & 204\% & 179\% & 163\% \\
      7 -- 8 & 162\% & 100\% & 162\% & 192\% & 160\% & 150\% \\
      \hline
      Overall & 181\% & 99\% & 181\% & 195\% & 193\% & 175\% \\
    \end{tabular}
  }
\end{table}

\subsection{Instruction-level Analysis of EVM Runtime}

\begin{figure}[H]
  \centering
  \includegraphics[width=80mm]{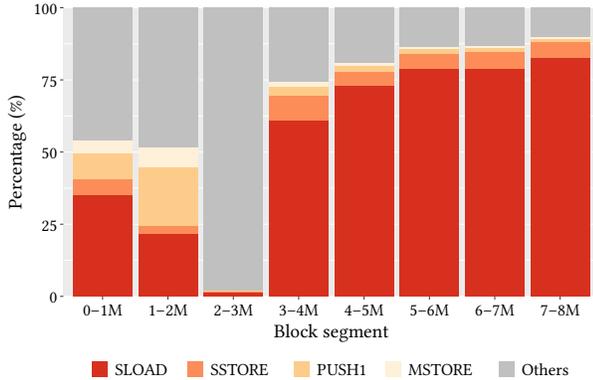}
  \caption{Overall execution time in percent.}
  \label{fig:inst-freq}
\end{figure}
Figure~\ref{fig:inst-freq} shows the percentage of execution time for the
most time-consuming instructions.
There was one irregularity when the Shanghai DoS attack occurred:
For the Shanghai DoS attack, the execution time is dominated by
\isrc{EXTCODESIZE} and \isrc{SUICIDE} instructions. Excluding that period, 
\isrc{SLOAD} appears as the most time-consuming
instruction. More importantly, the proportion of time 
spent on \isrc{SLOAD} continues to increase after Block \SI{3}{\BIGM} and
exceeds 80\% after Block \SI{7}{\BIGM}.


\begin{table}[H]
\small
  \caption{Average execution time of instructions (ns).\label{table:average-time-inst}}
  \resizebox{\columnwidth}{!} {
    \begin{tabular}{c|cccc}
Blocks (M) & SLOAD & SSTORE & PUSH1 & MSTORE \\
\hline
0 -- 1 &  5738  &  3751  &  85.4  &  158.5 \\
1 -- 2 &  8367  &  5844  &  79.2  &  107.5 \\
2 -- 3 &  8254  &  7025  &  92.2  &  224.1 \\
3 -- 4 &  18893  &  9646  &  94.3  &  214.4 \\
4 -- 5 &  37951  &  8130  &  85.9  &  175.6  \\
5 -- 6 &  51847  &  11512  &  79.6  &  157.7 \\
6 -- 7 &  68499  &  18952  &  80.7  &  149.0 \\
7 -- 8 &  82265 &  21480 &  78.2 &  153.9 \\
    \end{tabular}
  }
\end{table}

Table~\ref{table:average-time-inst} illustrates the average execution times of four major instructions
from Figure~\ref{fig:inst-freq}: \isrc{SLOAD}, \isrc{SSTORE}, \isrc{PUSH1}, and \isrc{MSTORE}\@.
We observe that the average execution times of \isrc{SLOAD} and \isrc{SSTORE} significantly increase as the blockchain grows.
Based on the fact that some instructions exhibit an increasing execution time as
the block-height increases, 
we divided instructions into two groups: block-height dependent~(BH-dependent)
and block-height independent~(BH-independent) instructions.

In this paper, we define an instruction as BH-dependent if it becomes slower when block-height increases,
otherwise it is a BH-independent instruction.
The correlation was calculated between block-height and average execution time of each instruction.
The instructions \isrc{SLOAD}, \isrc{SSTORE}, and \isrc{CALLCODE} were found to be BH-dependent instructions due to
their high positive correlation (larger than 0.7).

\subsection{Correspondence Between Macroscopic and Microscopic Instrumentation}
Although the instruction-level measurement is informative, the result may not match the
measurements of the macroscopic instrumentation. 
There could be super- or sub-linear runtime dependencies between subsequent EVM instructions.
In Figure~\ref{fig:relative-error}, we calculated the relative difference between the time measurements 
of the microscopic and the macroscopic
instrumentations. 
The relative difference is calculated as $\frac{\text{macro} - \text{micro}}{\text{macro}}$. In our experiment,
the relative difference
never exceeds 10\%.
\begin{figure}[H]
  \centering
  \includegraphics[width=80mm]{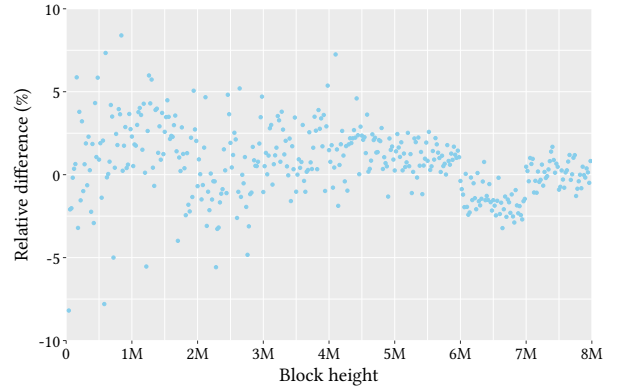}
  \caption{Relative difference between macroscopic and microscopic measurements.}
  \label{fig:relative-error}
\end{figure}
Differences are scattered in earlier blocks from 0 to \SI{2}{\BIGM}, because these
blocks contain only a minimal
number of transactions. 
Hence, the total execution time is small and the measurement is 
impacted by measurement noise.
After Block \SI{4}{\BIGM}, the number of transactions drastically increases as shown in Figure~\ref{fig:block-verification-plang9-ssd}, and
the relative differences become smaller ranging between $\pm$2.5\%.

By the central limit theorem~\cite{billingsley2008probability}, the measured
relative differences should follow a normal distribution. We conducted a Chi-square goodness
of fit test~\cite{cochran1952chi2} to check whether the data is
normally distributed.  The test resulted in $\chi_0^2 = 20.25$ with 17 degrees
of freedom. Because the critical value is 27.59 when $\alpha = 5\%$, the null
hypothesis that the data follows a normal distribution is accepted and the measurement is valid.

\subsection{Instrumentation Points in Parity and Geth\label{sec:geth:arch}}
Parity and Geth are the most popular Ethereum clients, being 
employed by more than 90\% of all Ethereum nodes
in the world~\cite{etherscan-nodetracker}.
\begin{table}[htpb]
\small
  \caption{Macro instrumentation points in Parity and Geth\label{tab:macro-points}}
  \resizebox{\columnwidth}{!} {
    \begin{tabular}{c|c}
 & Parity v2.4.0 \isrc{ethcore} module\\
\hline
Verify &  \isrc{verification::queue::VerificationQueue::verify} \\
Import &  \isrc{client::Client::import_verified_blocks} \\
DB &  \isrc{state::State::commit} \\
TX &  \isrc{state::State::execute} \\
EVM &  \isrc{evm::Interpreter::exec} \\
\hline
 & Geth v1.9.6 \isrc{core} module\\
\hline
Verify &  \isrc{BlockValidator.ValidateState} \\
Import &  \isrc{BlockChain.insertChain} \\
DB &  \isrc{state.StateDB.Finalise} \\
TX &  \isrc{ApplyMessage} \\
EVM &  \isrc{vm.run} \\
    \end{tabular}
  }
\end{table}
\begin{table}[htbp]
\small
  \caption{Micro instrumentation points in Parity and Geth\label{tab:micro-points}}
  \resizebox{\columnwidth}{!} {
    \begin{tabular}{c|c}
 & Parity v2.4.0 \isrc{ethcore} module\\
\hline
Message (CALL) & \isrc{executive::Executive::call} \\
Message (CREATE) & \isrc{executive::Executive::create} \\
EVM Execution & \isrc{evm::Interpreter::exec} \\
Instruction Verification & \isrc{evm::Interpreter::step_inner} \\
Instruction Execution & \isrc{evm::Interpreter::exec_instruction} \\
\hline
 & Geth v1.9.6 \isrc{core} module\\
\hline
Message (CALL) & \isrc{vm.EVM.Call} \\
Message (CREATE) & \isrc{vm.EVM.Create} \\
EVM Execution & \isrc{vm.run} \\
Instruction Verification & \isrc{vm.EVMInterpreter.Run} \\
Instruction Execution & \isrc{vm.operation.execute} \\
    \end{tabular}
  }
\end{table}
The software architecture of Geth is very similar to the software
architecture of Parity. Tables~\ref{tab:macro-points}~and~\ref{tab:micro-points} show
the correspondence of functional blocks between the two clients. 
We believe that the strong overlap stems from the Ethereum 
specification that dictates the transition functions for verification and
execution of blocks, transactions, and instructions in the EVM.
Hence,  our measurement methodology is not specialized for Parity,
but can be applied to other Ethereum clients.
We have very rudimentary data showing similar performance 
characteristics between Geth and Parity on the first five million 
blocks of the blockchain
on the macroscopic level.
\begin{center}
\begin{tabular}{rl|llll}
&0--\SI{1}{\BIGM}& 	1--\SI{2}{\BIGM}& 	2--\SI{3}{\BIGM}	& 3--\SI{4}{\BIGM}&	4--\SI{5}{\BIGM}\\
\hline 
Parity (macro.)	&17&	50&	384&	178&	1500\\
Geth (original)	&32&	71&	325&	222&	1381\\
\hline
\end{tabular}
\end{center}

\else
  \label{sec:geth:arch}
  \label{table:HWspec}
\fi 
\end{section}

\begin{section}{A Sustainable EVM Gas Cost Model\label{gas-model}}
\ifCOSTMODEL
In the previous section we showed that the execution time  of 
block-import increases with the block-height, and 
that the EVM takes the largest proportion of block-import in terms of runtime. 
However, it is not possible to deduce the economics of 
gas costs from these experiments.  In this section, we will 
introduce a model that analyzes execution time and gas cost of 
EVM instructions. With our experiment, we will show that the 
current gas metering in EVM is unsustainable. For this purpose,
we define the notion of a ``standard smart contract'' in the cost model. 
A standard smart contract simulates the performance behavior 
of an averaged program execution at an arbitrary block-height. 

\subsection{Modeling Execution Time and Gas on the Blockchain}

We introduce a model of execution for a ``standard`` smart contract.
For this purpose, we idealize the runtime computation of a 
smart contract. Node that the standard smart contract is a 
hypothetical smart-contract. It does not exist in the form of EVM 
bytecode; only in the cost model. 
It will reflect an average mix of instructions 
and an average execution length over the whole blockchain. 

The idealized model is defined as follows:
\begin{equation}
\text{avgprogtime}(n)=l_{p}\cdot\vec{t}^{T}(n)\cdot\vec{f_{p}},
\end{equation}
where $l_{p}$ is a scalar that represents the average execution length (i.e.\ number of instructions)
per smart contract invocation, time vector $\vec{t}(n)$ captures the runtime of EVM
instructions at block-height $n$, and frequency vector $f_{p}$ 
represents the normalized execution frequency
for the standard smart-contract execution,  i.e., $\sum\vec{f_{p}}^{T}\vec{1}=1$. 

The time vector $\vec{t}(n)$ and the execution frequency vector $\vec{f_{p}}$ 
have for each possible EVM instruction an element. The 
time vector $\vec{t}(n)$ is independent of a smart contract, and reflect
the execution time of a single instruction at a given block-height. 
For BH-independent instructions, we assume that the elements 
are constants. For BH-dependent instructions the corresponding 
elements will increase with a larger $n$.   The 
scalar $l_p$ and $\vec{f_p}$ depends on the standard 
smart contract $p$.   

In this idealized model, the execution of instructions is simulated in an additive fashion, 
i.e., every executed EVM instruction will contribute to the total runtime. 
This model is a coarse approximation of 
instruction execution because it does not take sub- or super-linear 
effectives between EVM instruction executions 
into accounts (i.e.\ that two dependent instructions either make the execution time longer
or shorter). Each instruction is considered in isolation.

For determining the parameters of the model, we can measure 
the time vector for each block/smart contract execution. The measurements 
can be later aggregated and normalized to deduce a time vector 
for the block chain. Similarly, $l_{p}$ and execution frequencies can be observed
for each program execution and aggregated later. From these 
estimates, we can deduce the performance characteristics of
a ``standard'' smart-contract. 

Similarly, we define an idealized gas cost for smart-contracts:
\begin{equation}
\text{avgproggas}(n)=l_{p}\cdot\vec{g}^{T}(n)\cdot\vec{f_{p}},
\end{equation}
where $g(n)$ is a vector containing for each instruction in the EVM an element.
There are some instructions with variant gas cost depending on the state of the program
including block number, values of input/output operands and account storage.
We used $\text{avgproggas}(n)$ in order to approximate gas cost of instructions
at different block-heights~$n$.

With the idealized execution time and gas model, we can define the ratio
between the two models 
\begin{equation}
\text{avgprogtpg}(n)=\frac{\text{avgprogtime}(n)}{\text{avgproggas}(n)}
\end{equation}
which we refer to as the average time per unit gas.

\subsection{Effects of Block-Height-Dependent Instructions}

\begin{figure}[H]
  \centering
  \includegraphics[width=80mm]{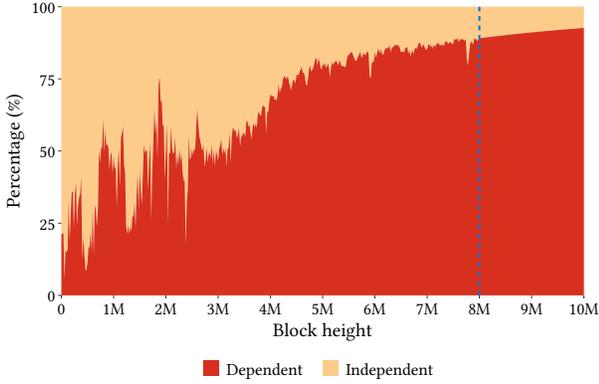}
  \caption{Execution time ratio of BH-dependent instructions and BH-independent instructions.
           Data beyond the block-height of 8\hspace{0.5mm}M is interpolated, as indicated by
           the dashed line.}
  \label{fig:dep-vs-indep}
\end{figure}

Figure~\ref{fig:dep-vs-indep} shows the execution time ratio between BH-dependent
and BH-independent instructions under the assumption that the ``standard'' contract executes 
at this block-height.  Note that the percentage of BH-dependent
instructions steadily increases and exceeds 80\% at block \num{8000000}, while the instruction mix
in the standard contract is constant - it is used as a yardstick. 

We also extrapolated the curve after \SI{8}{\million} blocks. To predict average execution time
of instructions after block \num{8000000},  we used average time for BH-independent instruction and
linear interpolation for BH-dependent ones. For BH-dependent instructions, we used polynomial
regression with one input parameter, block-height, and one output, average execution time of
the instruction. To avoid overfitting, we used randomly selected 80\% of observations for training
and the rest 20\% for validation. In the validation we used Bayesian information criterion
(BIC)~\cite{schwarz1978estimating} to evaluate polynomial models with different degrees.
With cross-validation by BIC, quadratic, cubic, and linear model were selected as the best
for \isrc{SLOAD}, \isrc{SSTORE}, and \isrc{CALLCODE}, respectively. The extrapolation 
is represented after blue dashed line in Figure~\ref{fig:dep-vs-indep}.
It shows that execution time ratio of BH-dependent instructions will keep increasing to
over 90\% at block \num{8500000}.

\subsection{New Gas Model for Constant Time-Per-Gas}

For sustainable Ethereum ecosystem, gas model must reflect real cost of computational resource
to the gas charged by smart contract execution.
Red lines in Figure~\ref{fig:gas-model} and Figure~\ref{fig:constant-time-per-gas} shows
the average gas cost and time per unit gas of the standard contract.
We observed that execution time of the standard contract increases but the gas cost remains the same.
As a result, time per unit gas is increasing over time (Figure~\ref{fig:constant-time-per-gas}).
This derives from the fact that BH-dependent instructions require more execution time in later blocks,
while their gas costs do not change. EIP-150~\cite{EIPs} was proposed to mitigate such irregularity between
gas costs and execution time of \isrc{SLOAD} and \isrc{SSTORE} in past, and it was accepted after block
\num{2463000} on the main network by EIP-608. However, EIP-150 only proposed constant gas costs for
\isrc{SLOAD} and \isrc{SSTORE}, thus the irregularity in BH-dependent instructions reappeared in later blocks.

We propose a new gas model to fix the increasing time per unit gas
by scaling the gas costs of BH-dependent instructions based on predicted execution time.
The goal of the new gas model is to stabilize time per unit gas to a constant $C$.
Let $t_i \left( n \right)$ be predicted execution time of instruction $i$ where $n$ is block-height.
For BH-dependent instruction $t_i \left( n \right)$ will be a polynomial function,
and for BH-independent one $t_i \left( n \right)$ will be a constant.
If $g_i \left( n \right)$ is a gas function of instruction $i$, the new gas model should
result as follows:
\begin{equation}
\frac{\Sigma t_i \left( n \right)}{\Sigma g_i\left(n\right)} = C
\end{equation} \label{eq:4}
To satisfy the equation above, we revised $g_i\left( n \right)$ of each instructions $i$ as follows:
\begin{equation}
g_i\left(n\right) = \frac{t_i \left( n \right)}{C}
\end{equation}

We defined $g_i\left( n \right)$ with our time prediction model that used to predict
$t_i\left( n \right)$ after \num{8000000} block in Figure~\ref{fig:dep-vs-indep}.
Thus, gas cost of BH-dependent instructions increases as blockchain grows, while
gas cost of the BH-independent stays constant.
We evaluated our new gas model with standard contract in Figure~\ref{fig:gas-model} and
Figure~\ref{fig:constant-time-per-gas}.
Red lines in Figure~\ref{fig:gas-model} shows that with current gas model, gas cost of
standard contract stays constant over blockchain growth, which leads time per unit gas of
standard contract depicted as red lines in Figure~\ref{fig:constant-time-per-gas} increases.
Blue lines in Figure~\ref{fig:gas-model} and Figure~\ref{fig:constant-time-per-gas} represent the result
of the new gas model. By increasing gas cost of BH-dependent instructions,
total gas cost of standard contract with our new model increases in Figure~\ref{fig:gas-model}.
Increasing gas cost from Figure~\ref{fig:gas-model} results time per unit gas of
standard contract stay constant in Figure~\ref{fig:constant-time-per-gas}.
In this paper we chose $C=5$ because time per unit gas in earlier blocks stays around 5.
\begin{figure}[H]
  \centering
  \includegraphics[width=80mm]{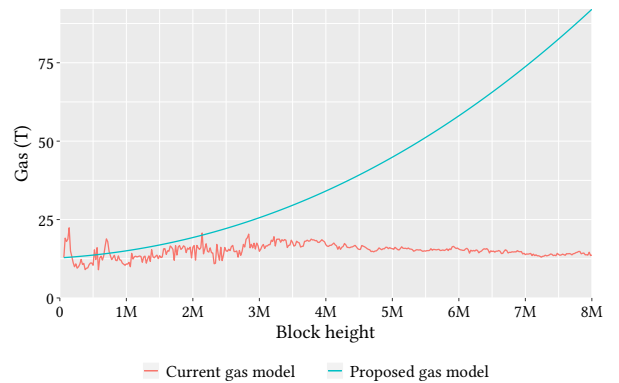}
  \caption{Comparison between current and proposed gas model.}
  \label{fig:gas-model}
\end{figure}

\begin{figure}[H]
  \centering
  \includegraphics[width=80mm]{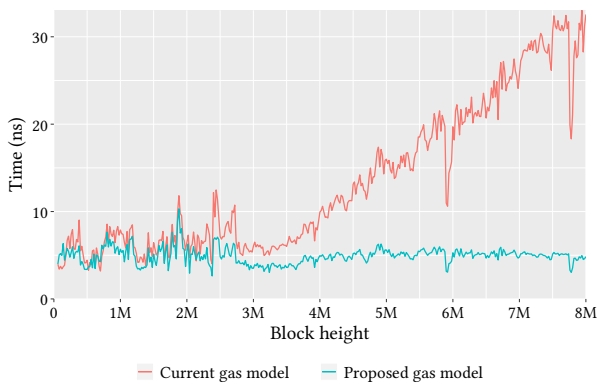}
  \caption{Time-per-gas of current and proposed gas model.}
  \label{fig:constant-time-per-gas}
\end{figure}

\subsubsection{Discussion of Gas cost Model}
The gas cost model was designed to equate the linearly increasing execution times
of blockchain size dependent instructions to have a linearly increasing gas cost.
Discrepancies between gas costs and execution times can lead to malicious users
sending transactions which use less gas but have higher execution time, thereby
allowing the attacker to abuse the resources of clients for minimal costs. The
new gas model will aims to reduce this attack vector by equating real execution
times with gas costs.

However, changing gas costs may cause some smart contracts, that rely on
specific gas costs of EVM instructions, to no longer function as designed.
Ethereum was designed for stability and immutability and the new gas model
provides a one-off-change that goes against immutability.
However, weighing the short term impact of making a
certain number of contracts unusable against the longer impact of increasing difference
between execution time and gas costs leading to potential DoS attacks and clients
re-allocating their resources, it is argued that the longer term benefits outweigh
the short term impacts.

Furthermore, while this gas cost model may add complexity to the calculations of
gas costs, it will allow programmers to pre-determine the gas costs of instructions
at certain block-heights. As a result, programmers will be able to design smart
contracts knowing whether their contract will still be usable going into the future.

An alternate method for implementing protection against the DoS attacks would be to
implement hard forks adjusting the cost of specific instructions. While this may
temporarily reduce the gap between execution time and gas costs, execution time will
continue to grow linearly and thus we will arrive back at the situation where
gas costs and execution time are not equated.

\fi 
\end{section}

\begin{section}{Related Work\label{sec:relwork}}
\subsection{Bitcoin Mining Analysis}
There has been a recent analysis directly on the costs of mining Bitcoin~\cite{bitc:carb}.
The paper shows an approximation to the amount of energy being spent on mining Bitcoin and
relates these findings to the energy consumptions of small countries.
While this research provides insight into the economy of mining in Bitcoin, it does not
cover the gas model of Ethereum. As Ethereum has the ability for state execution
which does not exist in Bitcoin it has its own economic model which is discussed in
this paper.

\subsection{Ethereum Technical Aspects}
High level concepts about the Ethereum network can be found through looking at
the Ethereum white paper, first release in 2013~\cite{buterin2013ethereum} by Vitalik
Buterin. For core technical details about Ethereum the yellow paper~\cite{Wood2014}
contains sufficient details to demonstrate the feasibility of the
concept.

For technical information specifically related to the Parity client the Parity
Ethereum Wiki~\cite{ParityWiki} gives a detailed
overview of the client, which is written in Rust programming language. It covers aspects
such as downloading a setting up a client, including package and dependency
requirements, configuration options and specifications on the recommended
hardware.

\subsection{Contract Bytecode \& Security}
A smart contract is stored on Ethereum as bytecode.
It is most commonly compiled from the Solidity programming language. Previous research
into contract bytecode tends to concentrate on security. Work from Data61~\cite{amani2018towards}
demonstrates the nature of conducting formal verification on Ethereum smart
contracts by segmenting the bytecode into basic blocks to confirm all possible
terminations of an execution. Static analysis tools such as Slither~\cite{slither}
and SmartCheck~\cite{tikhomirov2018smartcheck} have been developed to ananlyze
smart contracts for common flaws. Research generally in the form of blogs~\cite{gas-bytecode-blog}
has been done into the best ways to optimize gas when writing smart contracts.

\subsection{Cryptography}
Cryptography plays a key role in the Ethereum ecosystem. There are two main uses
of cryptography in the core blockchain and these are the hashing algorithm
Keccak256~\cite{bertoni2013keccak} and
a digital signature algorithm ECDSA~\cite{johnson2001elliptic}. Keccak is used as
part of the Ethash~\cite{EthashGithub} PoW algorithm to restrict the production
of blocks by enforcing the number of leading zero bits of the hash.
Transactions employ a digital ECDSA signature to prove a transaction
was authorized by the owner of the account. These cryptographic functions
are not the focus of this paper, however, monitoring their usage is.
Buterin~\cite{ethresearch-scaling} suggests an alternative signature scheme
that allows for the aggregation of signatures to significantly reduce verification
times.

\subsection{Ethereum Virtual Machine (EVM)}
The EVM is an interpreter that takes parameters from a user's
transaction and runs the bytecode stored by the network. The technical specifications of
the EVM can be found in the yellow paper~\cite{Wood2014}.
Currently there is research~\cite{ewasm-benches} being undertaken to compare the efficiency of the
EVM against another common web interpreter called WASM. The results of that research may
compliment the research undertaken in this paper as EVM may be constrained
resource and the benchmarks may show the areas which need improving. The Ethereum
network in the past has experience Denial Of Service attacks which exploited the
under pricing of gas with respect to instructions in the EVM. It is
important in understanding the efficiency of the client to understand what are the
inefficient instructions and how they can be exploited. Information about the
DoS attack and possible mitigation strategies are demonstrated in this paper~\cite{chen2017adaptive}.

\subsection{Scaling \& Bottlenecks}
The issue of scalability is heavily talked about in the Ethereum community and
is mentioned at inception in the yellow paper~\cite{Wood2014}. The core issue with
scalability is all nodes in the network are required to process all new blocks in order.
The current maximum transactions per second is about 15. Grayblock~\cite{grayblock-scaling}
describes how varying the block-time or gas limit (block size) will not solve the
issue, these variables may be able to increase the transactions per second
slightly but they will not be able to make a significant difference. To
make a significant difference two solutions can be applied, either make the
system parallelized as is being done in Ethereum2.0 or increase the efficiency
of the clients and block processing.

\subsection{Performance Measurements with Geth}
There exists measurement on instruction execution time on a virtual machine using
Geth~\cite{geth-measure}, which is measured on m5d.2xlarge instance of Amazon Web
Services. It also reports that storage operations are taking more time in
later blocks. Based on this observation, EIP-1884~\cite{EIPs} proposes to increase
gas of those instructions. However, the proposal is restricted that gas costs
will be increased to some constants, while execution time of the instructions will
keep growing by blockchain grows.

\subsection{Gas Mechanism and Resource Usage in EVM}
There are recent studies about irregularity of EVM gas mechanisms~\cite{perez2019broken,yang2019empirically}.
Yang et al.~\cite{yang2019empirically} vastly analyzed time-per-gas ratio of EVM instruction set
on both commodity and dedicated server hardware. It presented that the ratio is not uniform among
different instructions and some instructions have high variance in their own ratio due to state trie access.
\cite{perez2019broken} also reports varying gas for different instructions have low correlation with
their resource usage. It further investigated a relation between gas consumption and resource usages of
smart contracts, and concluded that storage usage is the most relevant and CPU usage is the least
relevant to gas consumption of smart contracts. However, both studies focused only on the misprized gas cost
itself and missed reasoning in which instructions exert most influence on current Ethereum gas model.
Moreover, they did little work on variability of instructions over growing blockchain
(e.g., \isrc{SLOAD} and \isrc{SSTORE}) and did not explore the effect from those instructions.


\subsection{Blockchain Databases}
Cohen, Rosenthal and Zohar~\cite{CohenBC:DB} explore the idea of using blockchains
as a storage layer for databases. The paper discusses possible methods for querying a
database that uses a blockchain for its storage layer. The article, however, does
not explore efficient solutions to deal with the expanding database size or how
to establish an effective cost model related to the size.
\end{section}

\begin{section}{Conclusion\label{sec:concl}}
Ethereum is the largest blockchain with the ability to execute
arbitrarily-expressive computations called smart contracts.
Users compensate miners for the execution of smart contracts.
We have shown that the costs of smart contract execution
are disproportionally larger than  the
computational costs, and that the cost-model of the underlying EVM instruction-set is wrongly modeled.
Our macroscopic instrumentation of the Parity Ethereum client
shows that both overall time and time per unit gas increase with block height. 
Our microscopic instrumentation revealed
that the execution time of three EVM instructions (\isrc{SLOAD}, \isrc{SSTORE}, and
\isrc{CALLCODE}) are drastically increasing with block-height, which is not covered
by the current Ethereum gas cost model. 
The Ethereum community is aware of the issue and
has increased gas costs sporadically by hard forks, which can only temporarily fix this issue.

Based on our performance data that we collected for the Ethereum blockchain up
to Block~\SI{8}{\BIGM}, we have devised a performance model to estimate gas
usage and execution time of a smart contract at a given block-height.  We have
proposed a new gas cost model that fixes the main irregularity of the current
Ethereum gas cost model. For the fix, we introduced the notion of a standard smart contract that 
simulates an average smart contract at an arbitrary block-height. 
Our new cost model stops the ongoing inflation of execution time per unit of gas.


\end{section}

\begin{acks}
%
%
This work was supported by
\grantsponsor{FANTOM}{Fantom Foundation}{https://fantom.foundation},
by the \grantsponsor{ARC}{Australian Government}{} through the ARC Discovery Project funding scheme
(\grantnum{ARC}{DP180104030}),
by the \grantsponsor{GRP}{General Research Program}{www.nrf.re.kr} through
the National Research Foundation of Korea (\grantnum{GRP}{NRF-2019R1F1A1062576}), and
by the \grantsponsor{NGICDP}{Next-Generation Information Computing Development Program}{www.nrf.re.kr}
through the National Research Foundation of Korea,
funded by the Ministry of Science, ICT \& Future Planning under
Grant No.~\grantnum{NGICDP}{NRF-2015M3C4A7065522}.
\end{acks}

\bibliographystyle{ACM-Reference-Format}
\bibliography{references}


\begin{thebibliography}{00}


\ifx \showCODEN    \undefined \def \showCODEN     #1{\unskip}     \fi
\ifx \showDOI      \undefined \def \showDOI       #1{{\tt DOI:}\penalty0{#1}\ }
  \fi
\ifx \showISBNx    \undefined \def \showISBNx     #1{\unskip}     \fi
\ifx \showISBNxiii \undefined \def \showISBNxiii  #1{\unskip}     \fi
\ifx \showISSN     \undefined \def \showISSN      #1{\unskip}     \fi
\ifx \showLCCN     \undefined \def \showLCCN      #1{\unskip}     \fi
\ifx \shownote     \undefined \def \shownote      #1{#1}          \fi
\ifx \showarticletitle \undefined \def \showarticletitle #1{#1}   \fi
\ifx \showURL      \undefined \def \showURL       #1{#1}          \fi
\providecommand\bibfield[2]{#2}
\providecommand\bibinfo[2]{#2}
\providecommand\natexlab[1]{#1}
\providecommand\showeprint[2][]{arXiv:#2}

\bibitem[\protect\citeauthoryear{2.0}{2.0}{2019a}]%
        {Ethereum:pos:cost}
\bibfield{author}{\bibinfo{person}{Ethereum 2.0}.}
  \bibinfo{year}{2018--2019}\natexlab{a}.
\newblock \bibinfo{title}{POS cost model}.
\newblock   (\bibinfo{year}{2018--2019}).
\newblock
\showURL{%
\url{https://docs.ethhub.io/ethereum-roadmap/ethereum-2.0/eth-2.0-economics/}}
\newblock
\shownote{(accessed 10/2019).}


\bibitem[\protect\citeauthoryear{2.0}{2.0}{2019b}]%
        {Ethereum:pos:transaction:fees}
\bibfield{author}{\bibinfo{person}{Ethereum 2.0}.}
  \bibinfo{year}{2018--2019}\natexlab{b}.
\newblock \bibinfo{title}{POS transaction fees}.
\newblock   (\bibinfo{year}{2018--2019}).
\newblock
\showURL{%
\url{https://github.com/ethereum/eth2.0-specs/blob/master/specs/core/0_beacon-chain.md\#transfers}}
\newblock
\shownote{(accessed 10/2019).}


\bibitem[\protect\citeauthoryear{Amani, B{\'e}gel, Bortin, and Staples}{Amani
  et~al\mbox{.}}{2018}]%
        {amani2018towards}
\bibfield{author}{\bibinfo{person}{Sidney Amani}, \bibinfo{person}{Myriam
  B{\'e}gel}, \bibinfo{person}{Maksym Bortin}, {and} \bibinfo{person}{Mark
  Staples}.} \bibinfo{year}{2018}\natexlab{}.
\newblock \showarticletitle{Towards verifying {Ethereum} smart contract
  bytecode in {I}sabelle/{HOL}}. In \bibinfo{booktitle}{{\em Proceedings of the
  7th ACM SIGPLAN International Conference on Certified Programs and Proofs}}.
  ACM, \bibinfo{pages}{66--77}.
\newblock


\bibitem[\protect\citeauthoryear{Amiri, Agrawal, and Abbadi}{Amiri
  et~al\mbox{.}}{2019}]%
        {caper}
\bibfield{author}{\bibinfo{person}{Mohammad~Javad Amiri},
  \bibinfo{person}{Divyakant Agrawal}, {and} \bibinfo{person}{Amr~El Abbadi}.}
  \bibinfo{year}{2019}\natexlab{}.
\newblock \showarticletitle{{CAPER}: A Cross-application Permissioned
  Blockchain}.
\newblock \bibinfo{journal}{{\em Proc. VLDB Endow.\/}} \bibinfo{volume}{12},
  \bibinfo{number}{11} (\bibinfo{date}{July} \bibinfo{year}{2019}),
  \bibinfo{pages}{1385--1398}.
\newblock
\showISSN{2150-8097}
\showDOI{%
\url{http://dx.doi.org/10.14778/3342263.3342275}}


\bibitem[\protect\citeauthoryear{Bertoni, Daemen, Peeters, and
  Van~Assche}{Bertoni et~al\mbox{.}}{2013}]%
        {bertoni2013keccak}
\bibfield{author}{\bibinfo{person}{Guido Bertoni}, \bibinfo{person}{Joan
  Daemen}, \bibinfo{person}{Micha{\"e}l Peeters}, {and} \bibinfo{person}{Gilles
  Van~Assche}.} \bibinfo{year}{2013}\natexlab{}.
\newblock \showarticletitle{Keccak}. In \bibinfo{booktitle}{{\em Annual
  international conference on the theory and applications of cryptographic
  techniques}}. Springer, \bibinfo{pages}{313--314}.
\newblock


\bibitem[\protect\citeauthoryear{Billingsley}{Billingsley}{2008}]%
        {billingsley2008probability}
\bibfield{author}{\bibinfo{person}{Patrick Billingsley}.}
  \bibinfo{year}{2008}\natexlab{}.
\newblock \bibinfo{booktitle}{{\em Probability and measure}}.
\newblock \bibinfo{publisher}{John Wiley \& Sons}.
\newblock


\bibitem[\protect\citeauthoryear{Boehm and Adve}{Boehm and Adve}{2008}]%
        {Boehm2008}
\bibfield{author}{\bibinfo{person}{Hans-J. Boehm} {and}
  \bibinfo{person}{Sarita~V. Adve}.} \bibinfo{year}{2008}\natexlab{}.
\newblock \showarticletitle{Foundations of the C++ Concurrency Memory Model}.
  In \bibinfo{booktitle}{{\em Proceedings of the 29th ACM SIGPLAN Conference on
  Programming Language Design and Implementation}} {\em (\bibinfo{series}{PLDI
  '08})}. \bibinfo{publisher}{ACM}, \bibinfo{address}{New York, NY, USA},
  \bibinfo{pages}{68--78}.
\newblock
\showISBNx{978-1-59593-860-2}
\showDOI{%
\url{http://dx.doi.org/10.1145/1375581.1375591}}


\bibitem[\protect\citeauthoryear{Buterin}{Buterin}{2019}]%
        {ethresearch-scaling}
\bibfield{author}{\bibinfo{person}{Vitalik Buterin}.}
  \bibinfo{year}{2018--2019}\natexlab{}.
\newblock \bibinfo{title}{On-chain scaling to potentially $\sim$500 tx/sec
  through mass tx validation}.
\newblock   (\bibinfo{year}{2018--2019}).
\newblock
\showURL{%
\url{https://ethresear.ch/t/on-chain-scaling-to-potentially-500-tx-sec-through-mass-tx-validation/3477}}
\newblock
\shownote{(accessed 08/2019).}


\bibitem[\protect\citeauthoryear{Buterin and Griffith}{Buterin and
  Griffith}{2017}]%
        {Buterin2017}
\bibfield{author}{\bibinfo{person}{Vitalik Buterin} {and}
  \bibinfo{person}{Virgil Griffith}.} \bibinfo{year}{2017}\natexlab{}.
\newblock \showarticletitle{Casper the Friendly Finality Gadget}.
\newblock \bibinfo{journal}{{\em CoRR\/}}  \bibinfo{volume}{abs/1710.09437}
  (\bibinfo{year}{2017}).
\newblock
\showURL{%
\url{http://arxiv.org/abs/1710.09437}}


\bibitem[\protect\citeauthoryear{Buterin, Reijsbergen, Leonardos, and
  Piliouras}{Buterin et~al\mbox{.}}{2019}]%
        {Casper:incentives}
\bibfield{author}{\bibinfo{person}{Vitalik Buterin},
  \bibinfo{person}{Dani{\"{e}}l Reijsbergen}, \bibinfo{person}{Stefanos
  Leonardos}, {and} \bibinfo{person}{Georgios Piliouras}.}
  \bibinfo{year}{2019}\natexlab{}.
\newblock \showarticletitle{Incentives in {E}thereum's Hybrid {C}asper
  Protocol}.
\newblock \bibinfo{journal}{{\em CoRR\/}}  \bibinfo{volume}{abs/1903.04205}
  (\bibinfo{year}{2019}).
\newblock
\showeprint[arxiv]{1903.04205}
\showURL{%
\url{http://arxiv.org/abs/1903.04205}}


\bibitem[\protect\citeauthoryear{Chen, Li, Wang, Chen, Li, Luo, Au, and
  Zhang}{Chen et~al\mbox{.}}{2017}]%
        {chen2017adaptive}
\bibfield{author}{\bibinfo{person}{Ting Chen}, \bibinfo{person}{Xiaoqi Li},
  \bibinfo{person}{Ying Wang}, \bibinfo{person}{Jiachi Chen},
  \bibinfo{person}{Zihao Li}, \bibinfo{person}{Xiapu Luo},
  \bibinfo{person}{Man~Ho Au}, {and} \bibinfo{person}{Xiaosong Zhang}.}
  \bibinfo{year}{2017}\natexlab{}.
\newblock \showarticletitle{An adaptive gas cost mechanism for Ethereum to
  defend against under-priced DoS attacks}. In \bibinfo{booktitle}{{\em
  International Conference on Information Security Practice and Experience}}.
  Springer, \bibinfo{pages}{3--24}.
\newblock


\bibitem[\protect\citeauthoryear{Cochran}{Cochran}{1952}]%
        {cochran1952chi2}
\bibfield{author}{\bibinfo{person}{William~Gemmell Cochran}.}
  \bibinfo{year}{1952}\natexlab{}.
\newblock \showarticletitle{The $\chi^2$ test of goodness of fit}.
\newblock \bibinfo{journal}{{\em The Annals of Mathematical Statistics\/}}
  (\bibinfo{year}{1952}), \bibinfo{pages}{315--345}.
\newblock


\bibitem[\protect\citeauthoryear{Cohen, Rosenthal, and Zohar}{Cohen
  et~al\mbox{.}}{2019}]%
        {CohenBC:DB}
\bibfield{author}{\bibinfo{person}{Sara Cohen}, \bibinfo{person}{Adam
  Rosenthal}, {and} \bibinfo{person}{Aviv Zohar}.}
  \bibinfo{year}{2019}\natexlab{}.
\newblock \showarticletitle{Reasoning about the Future in Blockchain
  Databases}. In \bibinfo{booktitle}{{\em BCDL: Proceedings of the 1st workshop
  on blockchain and distributed ledger at VLDB'2019}}.
\newblock


\bibitem[\protect\citeauthoryear{CoinSchedule}{CoinSchedule}{2019}]%
        {ICOstats}
\bibfield{author}{\bibinfo{person}{CoinSchedule}.}
  \bibinfo{year}{2019}\natexlab{}.
\newblock \bibinfo{title}{Crypto Token Sales Market Statistics}.
\newblock   (\bibinfo{year}{2019}).
\newblock
\showURL{%
\url{https://www.coinschedule.com/stats.php}}
\newblock
\shownote{(accessed 10/2019).}


\bibitem[\protect\citeauthoryear{community}{community}{2019b}]%
        {buterin2013ethereum}
\bibfield{author}{\bibinfo{person}{Ethereum community}.}
  \bibinfo{year}{2013--2019}\natexlab{b}.
\newblock \bibinfo{title}{Ethereum white paper}.
\newblock   (\bibinfo{year}{2013--2019}).
\newblock
\showURL{%
\url{https://github.com/ethereum/wiki/wiki/White-Paper}}
\newblock
\shownote{(accessed 08/2019).}


\bibitem[\protect\citeauthoryear{community}{community}{2019a}]%
        {EIPs}
\bibfield{author}{\bibinfo{person}{Ethereum community}.}
  \bibinfo{year}{2015--2019}\natexlab{a}.
\newblock \bibinfo{title}{Ethereum Improvement Proposals}.
\newblock   (\bibinfo{year}{2015--2019}).
\newblock
\showURL{%
\url{https://eips.ethereum.org/}}
\newblock
\shownote{(accessed 09/2019).}


\bibitem[\protect\citeauthoryear{Ethereum}{Ethereum}{2019b}]%
        {Patricia-Tree}
\bibfield{author}{\bibinfo{person}{Ethereum}.}
  \bibinfo{year}{2014--2019}\natexlab{b}.
\newblock \bibinfo{title}{Patricia Tree}.
\newblock   (\bibinfo{year}{2014--2019}).
\newblock
\showURL{%
\url{https://github.com/ethereum/wiki/wiki/Patricia-Tree/}}
\newblock
\shownote{(accessed 10/2019).}


\bibitem[\protect\citeauthoryear{Ethereum}{Ethereum}{2019c}]%
        {RLP}
\bibfield{author}{\bibinfo{person}{Ethereum}.}
  \bibinfo{year}{2014--2019}\natexlab{c}.
\newblock \bibinfo{title}{RLP}.
\newblock   (\bibinfo{year}{2014--2019}).
\newblock
\showURL{%
\url{http://github.com/ethereum/wiki/wiki/RLP/}}
\newblock
\shownote{(accessed 10/2019).}


\bibitem[\protect\citeauthoryear{Ethereum}{Ethereum}{2018}]%
        {Ethereum-Benchmarks}
\bibfield{author}{\bibinfo{person}{Ethereum}.}
  \bibinfo{year}{2016--2018}\natexlab{}.
\newblock \bibinfo{title}{Benchmarks}.
\newblock   (\bibinfo{year}{2016--2018}).
\newblock
\showURL{%
\url{https://github.com/ethereum/wiki/wiki/Benchmarks/}}
\newblock
\shownote{(accessed 10/2019).}


\bibitem[\protect\citeauthoryear{Ethereum}{Ethereum}{2019a}]%
        {Ethereum:pos:issuance}
\bibfield{author}{\bibinfo{person}{Ethereum}.}
  \bibinfo{year}{2018--2019}\natexlab{a}.
\newblock \bibinfo{title}{Monetary Policy}.
\newblock   (\bibinfo{year}{2018--2019}).
\newblock
\showURL{%
\url{https://docs.ethhub.io/ethereum-basics/monetary-policy/}}
\newblock
\shownote{(accessed 10/2019).}


\bibitem[\protect\citeauthoryear{Ethereum}{Ethereum}{2019d}]%
        {Geth}
\bibfield{author}{\bibinfo{person}{Go Ethereum}.}
  \bibinfo{year}{2013--2019}\natexlab{d}.
\newblock \bibinfo{title}{Go Ethereum}.
\newblock   (\bibinfo{year}{2013--2019}).
\newblock
\showURL{%
\url{https://geth.ethereum.org/}}
\newblock
\shownote{(accessed 08/2019).}


\bibitem[\protect\citeauthoryear{Etherscan}{Etherscan}{2019a}]%
        {etherscan-etherprice}
\bibfield{author}{\bibinfo{person}{Etherscan}.}
  \bibinfo{year}{2019}\natexlab{a}.
\newblock \bibinfo{title}{Ether Price History (USD)}.
\newblock   (\bibinfo{year}{2019}).
\newblock
\showURL{%
\url{https://etherscan.io/chart/etherprice}}
\newblock
\shownote{(accessed 10/2019).}


\bibitem[\protect\citeauthoryear{Etherscan}{Etherscan}{2019b}]%
        {etherscan-marketcap}
\bibfield{author}{\bibinfo{person}{Etherscan}.}
  \bibinfo{year}{2019}\natexlab{b}.
\newblock \bibinfo{title}{Ethereum Market Capitalization}.
\newblock   (\bibinfo{year}{2019}).
\newblock
\showURL{%
\url{https://etherscan.io/chart/marketcap}}
\newblock
\shownote{(accessed 15/10/2019).}


\bibitem[\protect\citeauthoryear{Etherscan}{Etherscan}{2019c}]%
        {etherscan-nodetracker}
\bibfield{author}{\bibinfo{person}{Etherscan}.}
  \bibinfo{year}{2019}\natexlab{c}.
\newblock \bibinfo{title}{Ethereum Node Tracker}.
\newblock   (\bibinfo{year}{2019}).
\newblock
\showURL{%
\url{https://etherscan.io/nodetracker}}
\newblock
\shownote{(accessed 10/2019).}


\bibitem[\protect\citeauthoryear{Flood and Goodenough}{Flood and
  Goodenough}{2015}]%
        {Flood2015}
\bibfield{author}{\bibinfo{person}{Mark~D Flood} {and}
  \bibinfo{person}{Oliver~R Goodenough}.} \bibinfo{year}{2015}\natexlab{}.
\newblock \bibinfo{title}{Contract as automaton: the computational
  representation of financial agreements}.
\newblock
  \bibinfo{howpublished}{\url{https://dx.doi.org/10.2139/ssrn.2648460}}.
  (\bibinfo{year}{2015}).
\newblock


\bibitem[\protect\citeauthoryear{Grayblock}{Grayblock}{2018}]%
        {grayblock-scaling}
\bibfield{author}{\bibinfo{person}{Grayblock}.}
  \bibinfo{year}{2018}\natexlab{}.
\newblock \bibinfo{title}{Blockchain Scaling}.
\newblock   (\bibinfo{year}{2018}).
\newblock
\showURL{%
\url{https://medium.com/coinmonks/blockchain-scaling-30c9e1b7db1b}}
\newblock
\shownote{(accessed 08/2019).}


\bibitem[\protect\citeauthoryear{Grech, Brent, Scholz, and Smaragdakis}{Grech
  et~al\mbox{.}}{2019}]%
        {icse19}
\bibfield{author}{\bibinfo{person}{Neville Grech}, \bibinfo{person}{Lexi
  Brent}, \bibinfo{person}{Bernhard Scholz}, {and} \bibinfo{person}{Yannis
  Smaragdakis}.} \bibinfo{year}{2019}\natexlab{}.
\newblock \showarticletitle{Gigahorse: thorough, declarative decompilation of
  smart contracts}. In \bibinfo{booktitle}{{\em Proceedings of the 41st
  International Conference on Software Engineering, {ICSE} 2019, Montreal, QC,
  Canada, May 25-31, 2019}}, \bibfield{editor}{\bibinfo{person}{Joanne~M.
  Atlee}, \bibinfo{person}{Tevfik Bultan}, {and} \bibinfo{person}{Jon Whittle}}
  (Eds.). \bibinfo{publisher}{{IEEE} / {ACM}}, \bibinfo{pages}{1176--1186}.
\newblock
\showISBNx{978-1-7281-0869-8}
\showDOI{%
\url{http://dx.doi.org/10.1109/ICSE.2019.00120}}


\bibitem[\protect\citeauthoryear{Gupta}{Gupta}{2019}]%
        {gas-bytecode-blog}
\bibfield{author}{\bibinfo{person}{Gupta}.} \bibinfo{year}{2019}\natexlab{}.
\newblock \bibinfo{title}{Solidity tips and tricks to save gas and reduce
  bytecode size}.
\newblock   (\bibinfo{year}{2019}).
\newblock
\showURL{%
\url{https://blog.polymath.network/solidity-tips-and-tricks-to-save-gas-and-reduce-bytecode-size-c44580b218e6}}
\newblock
\shownote{(accessed 08/2019).}


\bibitem[\protect\citeauthoryear{{Jack du Rose}}{{Jack du Rose}}{2017}]%
        {Rose17icos}
\bibfield{author}{\bibinfo{person}{{Jack du Rose}}.}
  \bibinfo{year}{2017}\natexlab{}.
\newblock \bibinfo{title}{Ethereum startups don't need Silicon Valley}.
\newblock   (\bibinfo{year}{2017}).
\newblock
\showURL{%
\url{https://venturebeat.com/2017/07/01/ethereum-startups-dont-need-silicon-valley/}}
\newblock
\shownote{(accessed 10/2019).}


\bibitem[\protect\citeauthoryear{Jakobsson and Juels}{Jakobsson and
  Juels}{1999}]%
        {Jakobsson:1999:POW}
\bibfield{author}{\bibinfo{person}{Markus Jakobsson} {and} \bibinfo{person}{Ari
  Juels}.} \bibinfo{year}{1999}\natexlab{}.
\newblock \showarticletitle{Proofs of work and bread pudding protocols}.
\newblock In \bibinfo{booktitle}{{\em Secure Information Networks}}.
  \bibinfo{publisher}{Springer}, \bibinfo{pages}{258--272}.
\newblock


\bibitem[\protect\citeauthoryear{Johnson, Menezes, and Vanstone}{Johnson
  et~al\mbox{.}}{2001}]%
        {johnson2001elliptic}
\bibfield{author}{\bibinfo{person}{Don Johnson}, \bibinfo{person}{Alfred
  Menezes}, {and} \bibinfo{person}{Scott Vanstone}.}
  \bibinfo{year}{2001}\natexlab{}.
\newblock \showarticletitle{The elliptic curve digital signature algorithm
  (ECDSA)}.
\newblock \bibinfo{journal}{{\em International journal of Information
  security\/}} \bibinfo{volume}{1}, \bibinfo{number}{1} (\bibinfo{year}{2001}),
  \bibinfo{pages}{36--63}.
\newblock


\bibitem[\protect\citeauthoryear{Kim, Ma, Murali, Mason, Miller, and
  Bailey}{Kim et~al\mbox{.}}{2018}]%
        {kim2018measuring}
\bibfield{author}{\bibinfo{person}{Seoung~Kyun Kim}, \bibinfo{person}{Zane Ma},
  \bibinfo{person}{Siddharth Murali}, \bibinfo{person}{Joshua Mason},
  \bibinfo{person}{Andrew Miller}, {and} \bibinfo{person}{Michael Bailey}.}
  \bibinfo{year}{2018}\natexlab{}.
\newblock \showarticletitle{Measuring {Ethereum} network peers}. In
  \bibinfo{booktitle}{{\em Proceedings of the Internet Measurement Conference
  2018}}. ACM, \bibinfo{pages}{91--104}.
\newblock


\bibitem[\protect\citeauthoryear{Nathan, Govindarajan, Saraf, Sethi, and
  Jayachandran}{Nathan et~al\mbox{.}}{2019}]%
        {BC:DB}
\bibfield{author}{\bibinfo{person}{Senthil Nathan}, \bibinfo{person}{Chander
  Govindarajan}, \bibinfo{person}{Adarsh Saraf}, \bibinfo{person}{Manish
  Sethi}, {and} \bibinfo{person}{Praveen Jayachandran}.}
  \bibinfo{year}{2019}\natexlab{}.
\newblock \showarticletitle{Blockchain Meets Database: Design and
  Implementation of a Blockchain Relational Database}.
\newblock \bibinfo{journal}{{\em Proc. VLDB Endow.\/}} \bibinfo{volume}{12},
  \bibinfo{number}{11} (\bibinfo{date}{July} \bibinfo{year}{2019}),
  \bibinfo{pages}{1539--1552}.
\newblock
\showISSN{2150-8097}
\showDOI{%
\url{http://dx.doi.org/10.14778/3342263.3342632}}


\bibitem[\protect\citeauthoryear{page}{page}{2018}]%
        {EthashGithub}
\bibfield{author}{\bibinfo{person}{Ethash~GitHub page}.}
  \bibinfo{year}{2015--2018}\natexlab{}.
\newblock \bibinfo{title}{Ethash}.
\newblock   (\bibinfo{year}{2015--2018}).
\newblock
\showURL{%
\url{https://github.com/ethereum/wiki/wiki/Ethash}}
\newblock
\shownote{(accessed 08/2019).}


\bibitem[\protect\citeauthoryear{page}{page}{2019a}]%
        {ParityGithub}
\bibfield{author}{\bibinfo{person}{Parity~GitHub page}.}
  \bibinfo{year}{2015--2019}\natexlab{a}.
\newblock \bibinfo{title}{Parity Ethereum 2.4.0}.
\newblock   (\bibinfo{year}{2015--2019}).
\newblock
\showURL{%
\url{https://github.com/paritytech/parity-ethereum/releases/tag/v2.4.0}}
\newblock
\shownote{(accessed 08/2019).}


\bibitem[\protect\citeauthoryear{page}{page}{2019b}]%
        {slither}
\bibfield{author}{\bibinfo{person}{Slither~Github page}.}
  \bibinfo{year}{2018--2019}\natexlab{b}.
\newblock \bibinfo{title}{Slither, the Solidity source analyzer}.
\newblock   (\bibinfo{year}{2018--2019}).
\newblock
\showURL{%
\url{https://github.com/crytic/slither}}
\newblock
\shownote{(accessed 08/2019).}


\bibitem[\protect\citeauthoryear{Perez and Livshits}{Perez and
  Livshits}{2019}]%
        {perez2019broken}
\bibfield{author}{\bibinfo{person}{Daniel Perez} {and}
  \bibinfo{person}{Benjamin Livshits}.} \bibinfo{year}{2019}\natexlab{}.
\newblock \showarticletitle{Broken Metre: Attacking Resource Metering in EVM}.
\newblock \bibinfo{journal}{{\em arXiv preprint arXiv:1909.07220\/}}
  (\bibinfo{year}{2019}).
\newblock


\bibitem[\protect\citeauthoryear{Pilkington}{Pilkington}{2016}]%
        {pilkington201611}
\bibfield{author}{\bibinfo{person}{Marc Pilkington}.}
  \bibinfo{year}{2016}\natexlab{}.
\newblock \showarticletitle{11 Blockchain technology: principles and
  applications}.
\newblock \bibinfo{journal}{{\em Research handbook on digital
  transformations\/}}  \bibinfo{volume}{225} (\bibinfo{year}{2016}).
\newblock


\bibitem[\protect\citeauthoryear{{R3 Consortium}}{{R3 Consortium}}{2019}]%
        {Corda2017}
\bibfield{author}{\bibinfo{person}{{R3 Consortium}}.}
  \bibinfo{year}{2016--2019}\natexlab{}.
\newblock \bibinfo{title}{Corda --- an open-source distributed ledger
  platform}.
\newblock   (\bibinfo{year}{2016--2019}).
\newblock
\showURL{%
\url{https://www.corda.net/}}
\newblock
\shownote{(accessed 10/2019).}


\bibitem[\protect\citeauthoryear{Schwarz}{Schwarz}{1978}]%
        {schwarz1978estimating}
\bibfield{author}{\bibinfo{person}{Gideon Schwarz}.}
  \bibinfo{year}{1978}\natexlab{}.
\newblock \showarticletitle{Estimating the dimension of a model}.
\newblock \bibinfo{journal}{{\em The annals of statistics\/}}
  \bibinfo{volume}{6}, \bibinfo{number}{2} (\bibinfo{year}{1978}),
  \bibinfo{pages}{461--464}.
\newblock


\bibitem[\protect\citeauthoryear{Source}{Source}{2019}]%
        {RocksDB}
\bibfield{author}{\bibinfo{person}{Facebook~Open Source}.}
  \bibinfo{year}{2012--2019}\natexlab{}.
\newblock \bibinfo{title}{RocksDB}.
\newblock   (\bibinfo{year}{2012--2019}).
\newblock
\showURL{%
\url{https://rocksdb.org/}}
\newblock
\shownote{(accessed 10/2019).}


\bibitem[\protect\citeauthoryear{Stoll, Klaaßen, and Gallersdörfer}{Stoll
  et~al\mbox{.}}{2019}]%
        {bitc:carb}
\bibfield{author}{\bibinfo{person}{Christian Stoll}, \bibinfo{person}{Lena
  Klaaßen}, {and} \bibinfo{person}{Ulrich Gallersdörfer}.}
  \bibinfo{year}{2019}\natexlab{}.
\newblock \showarticletitle{The Carbon Footprint of {B}itcoin}.
\newblock \bibinfo{journal}{{\em Joule\/}} (\bibinfo{year}{2019}).
\newblock


\bibitem[\protect\citeauthoryear{Swende}{Swende}{2019}]%
        {geth-measure}
\bibfield{author}{\bibinfo{person}{Martin~Holst Swende}.}
  \bibinfo{year}{2019}\natexlab{}.
\newblock \bibinfo{title}{VMstats}.
\newblock   (\bibinfo{year}{2019}).
\newblock
\showURL{%
\url{https://github.com/holiman/vmstats}}
\newblock
\shownote{(accessed 08/2019).}


\bibitem[\protect\citeauthoryear{Tapscott and Tapscott}{Tapscott and
  Tapscott}{2016}]%
        {Tapscott2016}
\bibfield{author}{\bibinfo{person}{Don Tapscott} {and} \bibinfo{person}{Alex
  Tapscott}.} \bibinfo{year}{2016}\natexlab{}.
\newblock \bibinfo{booktitle}{{\em Blockchain Revolution: How the Technology
  Behind Bitcoin Is Changing Money, Business, and the World}}.
\newblock \bibinfo{publisher}{Penguin}, \bibinfo{address}{London : Portfolio
  Penguin, 2016}.
\newblock
\showISBNx{9780241237854}


\bibitem[\protect\citeauthoryear{team}{team}{2019}]%
        {ewasm-benches}
\bibfield{author}{\bibinfo{person}{Ewasm team}.}
  \bibinfo{year}{2018--2019}\natexlab{}.
\newblock \bibinfo{title}{Instructions for benchmarking Ewasm precompiles}.
\newblock   (\bibinfo{year}{2018--2019}).
\newblock
\showURL{%
\url{https://github.com/ewasm/benchmarking}}
\newblock
\shownote{(accessed 08/2019).}


\bibitem[\protect\citeauthoryear{Technologies}{Technologies}{2019}]%
        {ParityWiki}
\bibfield{author}{\bibinfo{person}{Parity Technologies}.}
  \bibinfo{year}{2015--2019}\natexlab{}.
\newblock \bibinfo{title}{Parity Ethereum Wiki}.
\newblock   (\bibinfo{year}{2015--2019}).
\newblock
\showURL{%
\url{https://wiki.parity.io/Parity-Ethereum}}
\newblock
\shownote{(accessed 08/2019).}


\bibitem[\protect\citeauthoryear{Tikhomirov, Voskresenskaya, Ivanitskiy,
  Takhaviev, Marchenko, and Alexandrov}{Tikhomirov et~al\mbox{.}}{2018}]%
        {tikhomirov2018smartcheck}
\bibfield{author}{\bibinfo{person}{Sergei Tikhomirov},
  \bibinfo{person}{Ekaterina Voskresenskaya}, \bibinfo{person}{Ivan
  Ivanitskiy}, \bibinfo{person}{Ramil Takhaviev}, \bibinfo{person}{Evgeny
  Marchenko}, {and} \bibinfo{person}{Yaroslav Alexandrov}.}
  \bibinfo{year}{2018}\natexlab{}.
\newblock \showarticletitle{Smartcheck: Static analysis of {E}thereum smart
  contracts}. In \bibinfo{booktitle}{{\em 2018 IEEE/ACM 1st International
  Workshop on Emerging Trends in Software Engineering for Blockchain
  (WETSEB)}}. IEEE, \bibinfo{pages}{9--16}.
\newblock


\bibitem[\protect\citeauthoryear{{United Kingdom Government Office for
  Science}}{{United Kingdom Government Office for Science}}{2016}]%
        {UK2016}
\bibfield{author}{\bibinfo{person}{{United Kingdom Government Office for
  Science}}.} \bibinfo{year}{2016}\natexlab{}.
\newblock \bibinfo{title}{Distributed ledger technology: beyond block chain}.
\newblock
  \bibinfo{howpublished}{\url{https://www.gov.uk/government/publications/distributed-ledger-technology-blackett-review}}.
    (\bibinfo{year}{2016}).
\newblock


\bibitem[\protect\citeauthoryear{{Various}}{{Various}}{2017}]%
        {serpent}
\bibfield{author}{\bibinfo{person}{{Various}}.}
  \bibinfo{year}{2014--2017}\natexlab{}.
\newblock \bibinfo{title}{{GitHub - ethereum/serpent: The Serpent Language}}.
\newblock   (\bibinfo{year}{2014--2017}).
\newblock
\showURL{%
\url{https://github.com/ethereum/serpent}}
\newblock
\shownote{(accessed 10/2019).}


\bibitem[\protect\citeauthoryear{{Various}}{{Various}}{2019a}]%
        {solidity}
\bibfield{author}{\bibinfo{person}{{Various}}.}
  \bibinfo{year}{2015--2019}\natexlab{a}.
\newblock \bibinfo{title}{{GitHub - ethereum/solidity: The Solidity
  Contract-Oriented Programming Language}}.
\newblock   (\bibinfo{year}{2015--2019}).
\newblock
\showURL{%
\url{https://github.com/ethereum/solidity}}
\newblock
\shownote{{(accessed: 10/2019)}.}


\bibitem[\protect\citeauthoryear{{Various}}{{Various}}{2019b}]%
        {vyper}
\bibfield{author}{\bibinfo{person}{{Various}}.}
  \bibinfo{year}{2016--2019}\natexlab{b}.
\newblock \bibinfo{title}{{GitHub - ethereum/vyper: Pythonic Smart Contract
  Language for the EVM}}.
\newblock   (\bibinfo{year}{2016--2019}).
\newblock
\showURL{%
\url{https://github.com/ethereum/vyper}}
\newblock
\shownote{(accessed 10/2019).}


\bibitem[\protect\citeauthoryear{Wieczner}{Wieczner}{2017}]%
        {Fortune:ICOs}
\bibfield{author}{\bibinfo{person}{Jen Wieczner}.}
  \bibinfo{year}{2017}\natexlab{}.
\newblock \bibinfo{title}{Cryptocurrency ICOs Are Making Bitcoin Startups
  Richer than VCs Ever Did}.
\newblock
  \bibinfo{howpublished}{\url{http://fortune.com/2017/07/28/bitcoin-cryptocurrency-ico/}}.
    (\bibinfo{year}{2017}).
\newblock
\newblock
\shownote{(accessed 10/2019).}


\bibitem[\protect\citeauthoryear{Wood}{Wood}{2014}]%
        {Wood2014}
\bibfield{author}{\bibinfo{person}{Gavin Wood}.}
  \bibinfo{year}{2014}\natexlab{}.
\newblock \showarticletitle{Ethereum: A secure decentralised generalised
  transaction ledger}.
\newblock \bibinfo{journal}{{\em Ethereum project yellow paper\/}}
  (\bibinfo{year}{2014}), \bibinfo{pages}{1--32}.
\newblock


\bibitem[\protect\citeauthoryear{Yang, Murray, Rimba, and Parampalli}{Yang
  et~al\mbox{.}}{2019}]%
        {yang2019empirically}
\bibfield{author}{\bibinfo{person}{Renlord Yang}, \bibinfo{person}{Toby
  Murray}, \bibinfo{person}{Paul Rimba}, {and} \bibinfo{person}{Udaya
  Parampalli}.} \bibinfo{year}{2019}\natexlab{}.
\newblock \showarticletitle{Empirically Analyzing Ethereum's Gas Mechanism}.
\newblock \bibinfo{journal}{{\em arXiv preprint arXiv:1905.00553\/}}
  (\bibinfo{year}{2019}).
\newblock


\end{thebibliography}

\end{document}